\documentclass{aa}

\usepackage[english]{babel}
\usepackage[a4paper]{geometry}

\usepackage{graphicx}
\usepackage{threeparttable}
\usepackage{amsmath}
\usepackage[colorlinks=true, allcolors=blue]{hyperref}
\usepackage{lipsum}
\usepackage{xcolor}
\usepackage{comment}

\newcommand{\nv}{\hat{\bf n}}
\newcommand{\dzsz}{{\tt d0s0}}
\newcommand{\doso}{{\tt d1s1}}
\newcommand{\dmsm}{{\tt dmsm}}
\newcommand{\dtsf}{{\tt d10s5}}

\title{The Simons Observatory: pipeline comparison and validation for large-scale B-modes}
\titlerunning{The Simons Observatory: primordial B-modes}

\author{\small 
Kevin Wolz\inst{1,2}~\thanks{email: K.~Wolz, kevin.wolz93@gmail.com}
\and
Susanna Azzoni\inst{3,4}
\and
Carlos Herv\'{i}as-Caimapo\inst{5,6}
\and
Josquin Errard\inst{7}
\and 
Nicoletta Krachmalnicoff\inst{1,2,8}
\and
David Alonso\inst{3}
\and
Carlo Baccigalupi\inst{1,2,8}
\and
Ant\'on Baleato Lizancos\inst{9,10}
\and
Michael L. Brown\inst{11}
\and
Erminia Calabrese\inst{12}
\and
Jens Chluba\inst{11}
\and
Jo Dunkley\inst{17,18}
\and
Giulio Fabbian\inst{12,13}
\and
Nicholas Galitzki\inst{14}
\and
Baptiste Jost\inst{7,15}
\and
Magdy Morshed\inst{7, 19}
\and
Federico Nati\inst{16}
}

\institute{\small
International School for Advanced Studies (SISSA), Via Bonomea 265, I-34136 Trieste, Italy \goodbreak
\and
National Institute for Nuclear Physics (INFN) -- Sezione di Trieste, Via Valerio 2, I-34127 Trieste, Italy \goodbreak
\and
Department of Physics, University of Oxford, Denys Wilkinson Building, Keble Road, Oxford OX1 3RH, United Kingdom
 \goodbreak
\and
Kavli Institute for the Physics and Mathematics of the Universe (Kavli IPMU, WPI), UTIAS, The University of Tokyo, Kashiwa, Chiba 277-8583, Japan
\goodbreak
\and
Department of Physics, Florida State University, Tallahassee, Florida 32306, USA
\goodbreak
\and
Instituto de Astrof\'isica and Centro de Astro-Ingenier\'ia, Facultad de F\'isica, Pontificia Universidad Cat\'olica de Chile, Av. Vicu\~na Mackenna 4860, 7820436 Macul, Santiago, Chile
\goodbreak
\and
Universit\'{e} Paris Cit\'e, CNRS, Astroparticule et Cosmologie, F-75013 Paris, France
\goodbreak
\and
Institute for Fundamental Physics of the Universe (IFPU), Via Beirut 2, I-34151 Grignano (TS), Italy
\goodbreak
\and
Berkeley Center for Cosmological Physics, Department of Physics, University of California, Berkeley, CA 94720, USA
\goodbreak
\and
Lawrence Berkeley National Laboratory, One Cyclotron Road, Berkeley, CA 94720, USA
\goodbreak
\and
Jodrell Bank Centre for Astrophysics, School of Physics and Astronomy, The University of Manchester, Oxford Road, Manchester M20 4PE, United Kingdom
\goodbreak
\and
School of Physics and Astronomy, Cardiff University, The Parade, Cardiff, Wales CF24 3AA, UK
\goodbreak
\and
Center for Computational Astrophysics, Flatiron Institute, 162 5th Avenue, 10010, New York, NY, USA
\goodbreak
\and
Department of Physics, University of Texas at Austin, Austin, Texas 78722, USA
\goodbreak
\and
Kavli Institute for the Physics and Mathematics of the Universe (Kavli IPMU,WPI),  UTIAS,  The  University  of  Tokyo,  Kashiwa,  Chiba  277-8583, Japan
\goodbreak
\and
Department of Physics, University of Milano-Bicocca, Piazza della Scienza 3, 20126 Milan, Italy
\goodbreak
\and
Joseph Henry Laboratories of Physics, Jadwin Hall, Princeton University, Princeton, NJ, USA 08544
\goodbreak
\and
Department of Astrophysical Sciences, Peyton Hall, Princeton University, Princeton, NJ USA 08544
\goodbreak
\and
CNRS-UCB International Research Laboratory, Centre Pierre Binétruy, IRL2007, CPB-IN2P3, Berkeley, USA
}

\begin{document}

\abstract{The upcoming Simons Observatory Small Aperture Telescopes aim at achieving a constraint on the primordial tensor-to-scalar ratio $r$ at the level of $\sigma(r=0)\lesssim0.003$, observing the polarized CMB in the presence of partial sky coverage, cosmic variance, inhomogeneous non-white noise, and Galactic foregrounds.}
{We present three different analysis pipelines able to constrain $r$ given the latest available instrument performance, and compare their predictions on a set of sky simulations that allow us to explore a number of Galactic foreground models and elements of instrumental noise, relevant for the Simons Observatory.
}
{The three pipelines employ different combinations of parametric and non-parametric component separation at the map and power spectrum levels, and use B-mode purification to estimate the CMB B-mode power spectrum. We applied them to a common set of simulated realistic frequency maps, and compared and validated them with focus on their ability to extract robust constraints on the tensor-to-scalar ratio $r$. We evaluated their performance in terms of bias and statistical uncertainty on this parameter.
}
{In most of the scenarios the three methodologies achieve similar performance. Nevertheless, several simulations with complex foreground signals lead to a $>2\sigma$ bias on $r$ if analyzed with the default versions of these pipelines, highlighting the need for more sophisticated pipeline components that marginalize over foreground residuals. We show two such extensions, using power-spectrum-based and map-based methods, that are able to fully reduce the bias on $r$ below the statistical uncertainties in all foreground models explored, at a moderate cost in terms of $\sigma(r)$.}{}

\maketitle

\section{Introduction}\label{sec:intro}
  
   One of the next frontiers in cosmological science using the cosmic microwave background (CMB) is the observation of large-scale B-mode polarization, and the consequent potential detection of primordial gravitational waves. Such a detection would grant us a glance into the infant Universe and its high-energy physics, at scales unattainable by any other experiment. Primordial tensor perturbations, which would constitute a stochastic background of primordial gravitational waves, would source a parity-odd B-mode component in the polarization of the CMB \citep{1997PhRvL..78.2058K,1997ApJ...482....6S,1997PhRvL..78.2054S,1997PhRvD..55.1830Z}. The ratio between the amplitudes of the primordial power spectrum of these tensor perturbations and the primordial spectrum of the scalar perturbations is referred to as the tensor-to-scalar ratio $r$. This ratio covers a broad class of models of the early Universe, allowing us to test and discriminate between models that predict a wide range of values of $r$. These include vanishingly small values, as resulting from models of quantum gravity \citep[e.g.][]{2018CQGra..35m5004I,2019PhLB..795..666I}, as well as those expected to soon enter the detectable range, predicted by models of inflation \citep{1979JETPL..30..682S,1984NuPhB.244..541A,2014PDU.....5...75M, 2014JCAP...03..039M,2020A&A...641A..10P}. An unequivocal measurement of $r$, or a stringent upper bound, would thus greatly constrain the landscape of theories of the early Universe.
  
  Although there is no evidence of primordial B-modes yet, current CMB experiments place stringent constraints on their amplitude, finding $r < 0.036$ at 95\% confidence \citep{2021PhRvL.127o1301A} when evaluated at a pivot scale of 0.05 Mpc$^{-1}$. At the same time, these experiments firmly establish that the power spectrum of primordial scalar perturbations is not exactly scale-independent, with the scalar spectral index $n_s-1\sim0.03$ \citep[e.g.][]{2020A&A...641A...6P}. Given this measurement, several classes of inflationary models predict $r$ to be in the $\sim10^{-3}$ range \citep[see][and references therein]{2016ARA&A..54..227K}.

  Even though the only source of primordial large-scale B-modes at linear order are tensor fluctuations, in practice, a measurement is complicated by several factors: first, the gravitational deflection of the background CMB photons by the cosmic large-scale structure creates coherent sub-degree distortions in the CMB, known as CMB lensing \citep{2006PhR...429....1L}. Through this mechanism, the nonlinear scalar perturbations from the late Universe transform a fraction of the parity-even E-modes into B-modes at intermediate and small scales \citep{1998PhRvD..58b3003Z}. Second, diffuse Galactic foregrounds have significant polarized emission, and in particular foreground components such as synchrotron radiation and thermal emission from dust produce B-modes with a significant amplitude. Component separation methods, which exploit the different spectral energy distributions (SED) of the CMB and foregrounds to separate the different components, are thus of vital importance \citep{Delabrouille:2007bq,2008A&A...491..597L}. Practical implementations of these methods must also be able to carry out this separation in the presence of instrumental noise and systematic effects \citep[e.g.][]{2018JCAP...04..022N,2021JCAP...05..032A}.

  Polarized Galactic foregrounds pose a formidable obstacle when attempting to measure primordial B-modes at the level of $r \sim 10^{-3}$. Current measurements of Galactic emission demonstrate that at the relevant scales, the Galactic B-mode signal would dominate any existing primordial signal \citep{2016A&A...594A..10P,2016A&A...586A.133P,2020A&A...641A...4P,2020A&A...641A..11P}. At the minimum of polarized Galactic thermal dust and synchrotron emission, around 80\,GHz, their B-mode signal represents an effective tensor-to-scalar ratio with amplitude larger than the target CMB signal, even in the cleanest regions of the sky \citep{2016A&A...588A..65K}. Component separation methods are able to clean most of this, but small residuals left after the cleaning could be comparable to the primordial B-mode signal we want to measure. Many recent works analyze this problem and make forecasts on how well we could potentially measure $r$ with different ground-based and satellite experiments \citep[e.g.][]{2009A&A...503..691B,2011MNRAS.414..615B,2011ApJ...737...78K,2012MNRAS.424.1914A,2012PhRvD..85h3006E,2016MNRAS.458.2032R,2016PhRvD..94h3526S,2016JCAP...03..052E,2017MNRAS.468.4408H,2017PhRvD..95d3504A,2018MNRAS.474.3889R,2018JCAP...04..023R,Errard2019,2019arXiv190508888T,2021MNRAS.503.2478R,2021JCAP...05..047A,2022ApJ...924...11H,2022ApJ...926...54A,2022A&A...660A.111V,PTEP2022}. These works highlight that, if left untreated, systematic residuals from an overly simplistic characterization of foregrounds will bias an $r\sim10^{-3}$ measurement by several $\sigma$. Thus, it is of vital importance to model the required foreground complexity when cleaning the multi-frequency CMB observations, and to keep a tight control over systematics without introducing significant bias.
  
  Multiple upcoming CMB experiments rank the detection of large-scale primordial B-modes among their primary science targets. Near-future experiments such as the BICEP Array \citep{biceparr} target a detection at the level of $r \sim 0.01$, while in the following decade, next-generation projects, such as LiteBIRD \citep{2019JLTP..194..443H} and CMB-S4 \citep{2016arXiv161002743A}, will aim at $r \sim 0.001$.
  
  The Simons Observatory (SO), like the BICEP Array, targets the detection of primordial gravitational waves at the level of $r\sim0.01$ \citep[see ``The Simons Observatory: science goals and forecasts'',][]{2019JCAP...02..056A}, and its performance at realizing this goal is the main focus of this paper. SO is a ground-based experiment, located at the Cerro Toco site in the Chilean Atacama desert, which observes the microwave sky in six frequency channels, from 27 to 280\,GHz, with full science observations scheduled to start in 2024. SO consists of two main instruments. On the one hand, a Large Aperture Telescope (LAT) with a 6m diameter aperture targets small-scale CMB physics, secondary anisotropies, and the CMB lensing signal. Measurements of the latter will serve to subtract lensing-induced B-modes from the CMB signal to retrieve primordial B-modes \citep[using a technique  known as ``delensing'', see][]{2022PhRvD.105b3511N}. On the other hand, multiple Small Aperture Telescopes (SATs) with 0.4m diameter apertures will make large-scale, deep observations of $\sim 10$\% of the sky, with the main aim of constraining the primordial B-mode signal, peaking on scales $\ell \sim 80$ (the so-called ``recombination bump''). We refer to \cite{2019JCAP...02..056A} for an extended discussion on experimental capabilities.

  In this paper, we aim at validating three independent B-mode analysis pipelines. We compare their performance regarding a potential $r$ measurement by the SO SATs, and evaluate the capability of the survey to constrain $\sigma(r=0) \leq 0.003$ in the presence of foreground contamination and instrumental noise. To that end, we produce sky simulations encompassing different levels of foreground complexity, CMB with different values of $r$ and different amounts of residual lensing contamination, and various levels of the latest available instrumental noise\footnote{We note that the pipelines are still agnostic to some aspects of the instrumental noise such as filtering, which may impact the overall forecasted scientific performance. We anticipate studying these in detail in future work.}, calculated from the parametric models presented in \cite{2019JCAP...02..056A}.
   
  We feed these simulations through the analysis pipelines and test their performance, quantifying the bias and statistical uncertainty on $r$ as a function of foreground and noise complexity. The three pipelines are described in detail in Sect.~\ref{sec:pipelines}. Section~\ref{sec:sims} presents the simulations used in the analysis, including the models used to produce CMB and foreground sky maps, as well as instrumental noise. In Sect.~\ref{sec:results}, we present our forecasts for $r$, the power spectrum products, and a comparison of the relative weights assigned to the individual frequency channels when recovering the cleaned CMB. Section~\ref{sec:d10s5} shows preliminary results on a set of new, complex foreground simulations. In Sect.~\ref{sec:conc} we summarize and draw our conclusions. Appendix~\ref{app:validation} summarizes the $\chi^2$ analysis performed on the cross-$C_\ell$ cleaning pipeline, while Appendix~\ref{sec:nilc_gaussian_bias} discusses biases on Gaussian simulations observed with the NILC cleaning pipeline.

\section{Methods, pipelines} \label{sec:pipelines}

In this section we present our three component separation pipelines, that adopt complementary approaches widely used in the literature: power-spectrum-based parametric cleaning (\cite{2016PhRvL.116c1302B,2018PhRvL.121v1301B}), Needlet Internal Linear Combination (NILC) blind cleaning \citep[][]{2009A&A...493..835D,2012MNRAS.419.1163B,2013MNRAS.435...18B}, and map-based parametric cleaning \citep{Poletti_Errard_2022}. In the following, these are denominated pipelines A, B, and C, respectively. The cleaning algorithms operate on different data spaces (harmonic, needlet, and pixel space) and vary in their cleaning strategy (parametric, meaning that we assume an explicit model for the frequency spectrum of the foreground components, or blind, meaning that we do not model the foregrounds or make any assumptions on what their frequency spectrum should be). Hence, they do not share the same set of method-induced systematic errors. This will serve as an important argument in favor of claiming robustness of our inference results. 

Table~\ref{tab:pipelines_overview} lists the three pipelines and their main properties. Although there are some similarities between these analysis pipelines and the forecasting frameworks that were exploited in \cite{2019JCAP...02..056A}, the tools developed for this paper are novel implementations designed to deal with realistic SO data-like inputs, including complex noise and more exotic foreground simulations compared to what was considered in the previous work. We stress again that no filtering or other systematic effects were included in the noise maps.

\begin{table*}
\centering
\caption{Overview of the component separation pipelines used to infer $r$}\label{tab:pipelines_overview}
\begin{tabular}{ccccc}
\hline
\hline
\noalign{\smallskip}
Pipeline & method & data space & blind / parametric & $r$ inference step \\
\noalign{\smallskip}
\hline
\noalign{\smallskip}
A & Cross-$C_\ell$ cleaning & harmonic (power spectra) & parametric & multi-frequency $C_\ell$ likelihood\\
B & NILC cleaning & needlets (maps) & blind & CMB-only $C_\ell$ likelihood\\
C & map-based cleaning & pixels (maps) & parametric & CMB-only $C_\ell$ likelihood\\
\hline
\noalign{\smallskip}
\hline
\end{tabular}

\end{table*}

  \subsection{Pipeline A: Cross-$C_\ell$ cleaning}\label{ssec:pipelines.Cell}
\begin{figure}
\centering
 \includegraphics[width=\columnwidth]{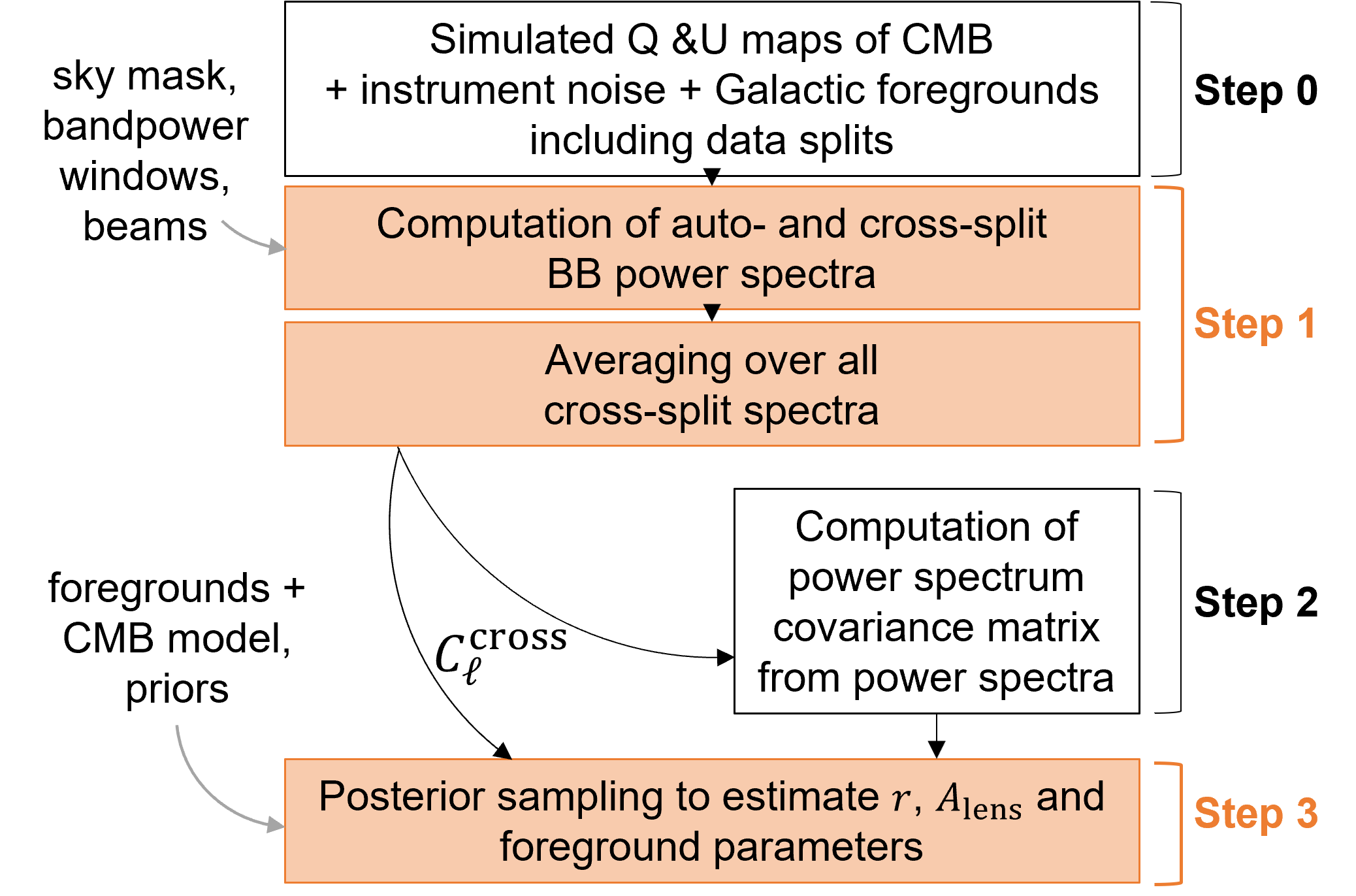}
\caption{Schematic of pipeline A. Orange colors mark steps that are repeated 500 times, once for each simulation.} \label{fig:schematic_a}
\end{figure}

    Pipeline A is based on a multi-frequency power-spectrum-based component separation method, similar to that used in the latest analysis carried out by the BICEP/{\sl Keck} collaboration \citep{2016PhRvL.116c1302B,2018PhRvL.121v1301B}. The data vector is the full set of cross-power spectra between all frequency maps, $C_\ell^{\nu\nu'}$. The likelihood compares this against a theoretical prediction that propagates the map-level sky and instrument model to the corresponding power spectra. The full pipeline is publicly available\footnote{See \href{https://github.com/simonsobs/BBPower}{github.com/simonsobs/BBPower}.}, and a schematic overview is provided in Fig.~\ref{fig:schematic_a}.

\begin{figure}
\centering
 \includegraphics[width=\columnwidth]{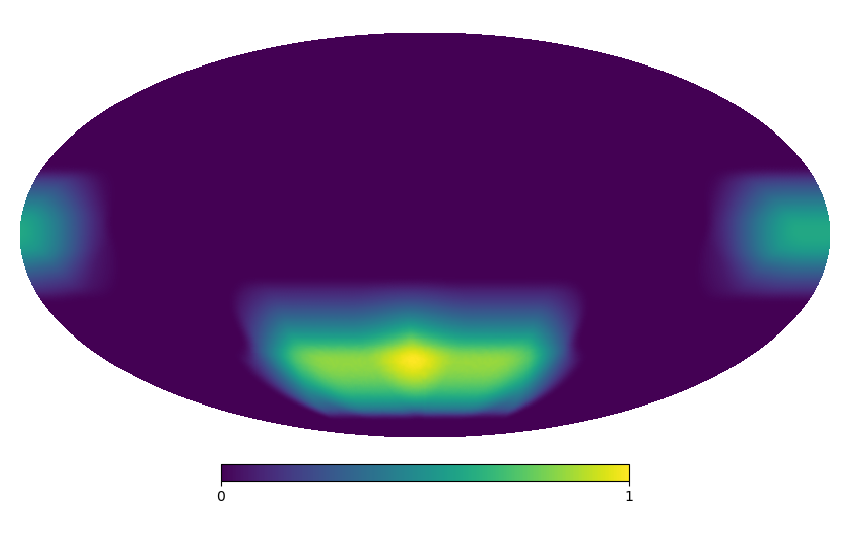}
\caption{Apodized SAT hits map with effective $f_{\rm sky}=10\%$ used in this paper, shown in equatorial projection. Mask edges are apodized using a C1-type kernel with an apodization scale of 10 degrees. } \label{fig:mask}
\end{figure}

    In step 1, power spectra are measured using a pseudo-$C_\ell$ approach with B-mode purification as implemented in {\tt NaMaster} \citep{2019MNRAS.484.4127A}, accounting for the leakage of E-mode power into B-mode power caused by incomplete sky coverage. As described in \cite{2007PhRvD..76d3001S} the presence of a sky mask leads to the presence of ambiguous modes contaminated by full-sky E-modes. These must be removed at the map level to avoid the contribution to the power spectrum uncertainties from the leaked E-modes. The mask used for this analysis traces the hits count map released in \cite{2019JCAP...02..056A} (see Fig.~\ref{fig:mask}), and its edges are apodized using a C1-type kernel \citep[see][]{2009PhRvD..79l3515G} with an apodization scale of 10 degrees, yielding an effective sky coverage of $f_{\rm sky}\sim10\%$. Each power spectrum is calculated in bandpower windows with constant bin width $\Delta\ell=10$, of which we only keep the range $30\leq\ell\leq300$. Our assumption is that on real data, larger scales are contaminated by atmospheric noise and filtering, whereas smaller scales, targeted by the SO-LAT and useful for constraining lensing B-modes, do not contain any significant primordial B-mode contribution. To avoid a significant bias in the auto-correlations when removing the impact of instrumental noise, a precise noise model is required that may not be available in practice. We address this issue by using data splits, which in the case of real data may be formed by subdividing data among observation periods, sets of detectors, sky patches, or by other means, while in this paper, we resort to simulations.
    
\begin{figure*}
\centering
 \includegraphics[width=\textwidth]{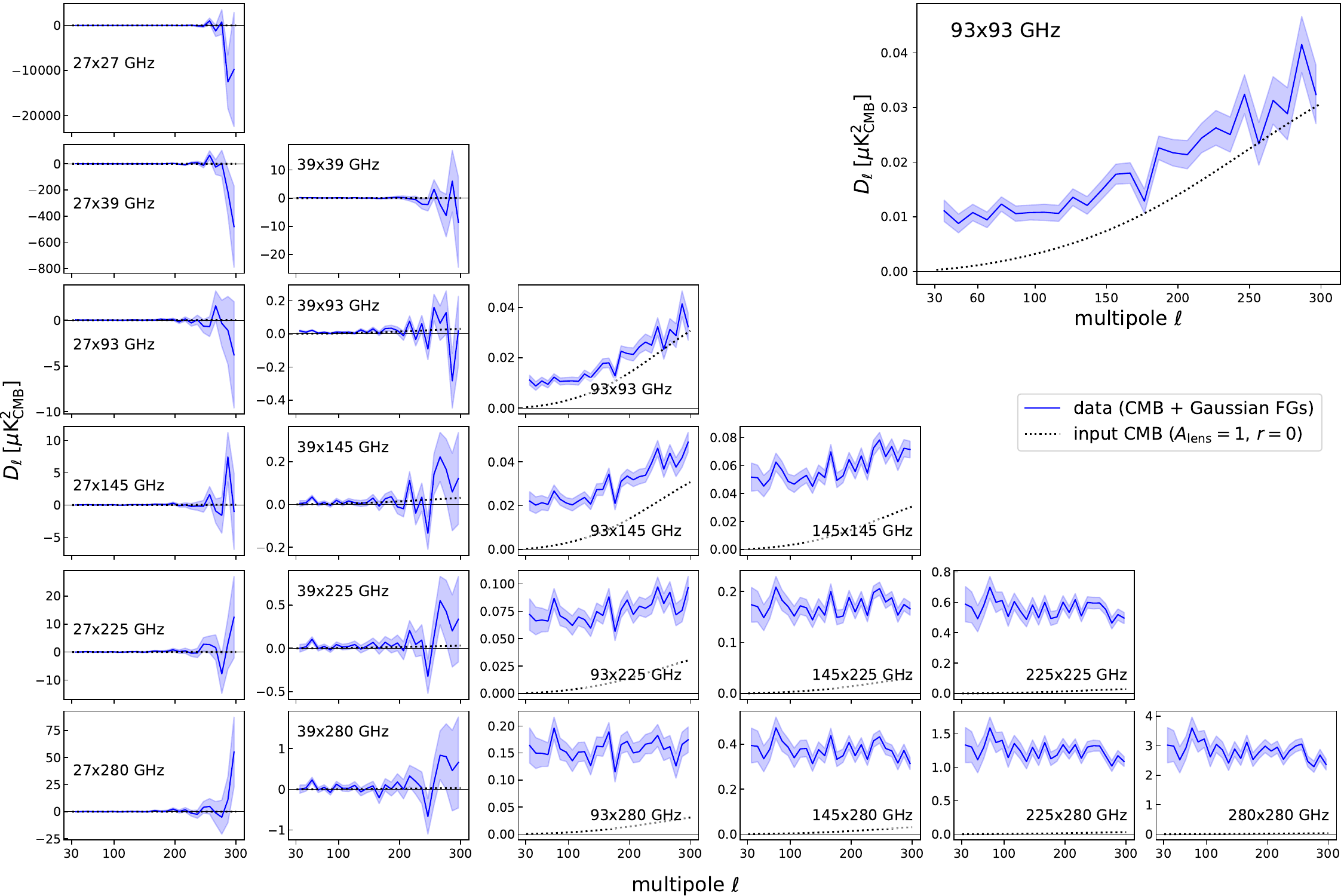}
\caption{Simulated power spectrum input data analyzed by pipeline A. We show a single realization of CMB and Gaussian foregrounds. Blue shaded areas quantify the 1$\sigma$ Gaussian standard deviation calculated from simulations of CMB, noise, and Gaussian foregrounds. We note that negative auto-spectra can occur at noise-dominated scales as a result of cross-correlating data splits.} \label{fig:raw_data_a}
\end{figure*}

    We construct simulated observations for each sky realization comprising $S=4$ independent splits with the same sky but different noise realizations (each with a commensurately larger noise amplitude). We compute $BB$ power spectra from pairs of maps, each associated with a given data split and a given frequency channel. For any fixed channel pair combination, we average over the corresponding set of $S(S-1)/2=6$ power spectra with unequal split pairings. For $N=6$ SAT frequency channels, this results in a collection of $N(N+1)/2=21$ noise-debiased multi-frequency power spectra, shown in Fig.~\ref{fig:raw_data_a}. We note that, in principle, we could model and subtract the noise bias explicitly, since we have full control over the noise properties in our simulations. In realistic settings, however, the accuracy of an assumed noise model may be limited. While inaccurate noise modeling would affect the statistical uncertainty $\sigma(r)$ through the covariance matrix calculated from simulations, the cross-split approach ensures robustness of the inferred value of $r$ against noise-induced bias.

    In step 2, we estimate the bandpower covariance matrix from simulations, assuming no correlations between different multipole windows. We note that, since our realistic foreground templates cannot be used as statistical samples, the covariance computation assumes Gaussian foreground simulations (see Sect.~\ref{sec:sims}) to include foreground signal variance in the budget. As we show in Appendix~\ref{app:validation}, this covariance matrix is indeed appropriate, as it leads to the theoretically expected empirical distribution of the $\hat{\chi}^2_{\rm min}$ statistic not only in the case of Gaussian foregrounds, but also for the non-Gaussian foreground simulations. Inaccurate covariance estimates would make this statistic peak at higher or lower values, which we do not observe.
    
    Step 3 is the parameter inference stage. We use a Gaussian likelihood when comparing the multi-frequency power spectra with their theoretical prediction. We note that, in general, the power spectrum likelihood is non-Gaussian, and \cite{hamimechelewis2008} provide an approximate likelihood that is able to account for this non-Gaussianity. We explicitly verified that both likelihoods lead to equivalent parameter constraints, and thus choose the simpler Gaussian option. The validity of the Gaussian approximation is a consequence of the central limit theorem, since each measured bandpower consists of effectively averaging over $N_{\rm modes}\simeq\Delta\ell\times f_{\rm{sky}}\times(2\ell+1) > 61$ independent squared modes on the scales used here. We note that this assumption is valid thanks to the relatively large SO-SAT sky patch and would not hold any longer for the smaller BICEP/{\sl Keck}-like patch sizes. The default sky model is the same as that described in \cite{2021JCAP...05..032A}. We model the angular power spectra of dust and synchrotron as power laws of the form $D_\ell=A_c(\ell/\ell_0)^{\alpha_c}$, with $\ell_0=80$, and $c=d$ or $s$ for dust and synchrotron, respectively. The dust SED is modeled as a modified black-body spectrum with spectral index $\beta_d$ and temperature $\Theta_d$, which we fix to $\Theta_d=20\,{\rm K}$. The synchrotron SED is modeled as a power law with spectral index $\beta_s$. Finally, we consider a dust-synchrotron correlation parameter $\epsilon_{ds}$. Including the tensor-to-scalar ratio $r$ and a free lensing B-mode amplitude $A_{\rm lens}$, this fiducial model has nine free parameters:
    \begin{equation}
    \{A_{\rm lens},r,A_d,\alpha_d,\beta_d,A_s,\alpha_s,\beta_s,\epsilon_{ds}\}.
    \end{equation}
    We refer to this method as ``$C_\ell$-fiducial''. Table~\ref{tab:priors_pipeline_a} lists the priors on its parameters. \\

    The main drawback of power-spectrum-based pipelines in their simplest incarnation, is their inability to account for spatial variation in the foreground spectra. If ignored, this spatial variation can lead to biases at the level of $r\sim O(10^{-3})$, which are significant for the SO target. At the power spectrum level, spatially-varying SEDs give rise to frequency decorrelation, which can be included in the model. In this work, we show results for an extended model that uses the moment-based   We refer to results using this method as ``$C_\ell$-moments'', or ``A + moments''. The priors in the shared parameter space are the same as for $C_\ell$-fiducial. Table~\ref{tab:priors_pipeline_a} lists the priors on its additional four parameters. For both methods, we sample posteriors using the \texttt{emcee} code \citep{2013PASP..125..306F}. It should be noted that we assume a top-hat prior on $r$ in the range $[-0.1,\, 0.1]$ instead of imposing $r>0$. The reason is that we would like to remain sensitive to potential negative biases on $r$. While negative $r$ values do not make sense physically, they may result from volume effects caused by choosing specific priors on other parameters that we marginalize over. Opening the prior on $r$ to negative values allows us to monitor these unwanted effects, offering a simple robustness check. On real data, this will be replaced by a positivity prior $r>0$, but only after ensuring that our specific prior choices on the other parameters do not bias $r$, which is the focus of future work.
  
\begin{table*}
\centering
\caption{Parameter priors for pipeline A, considering both the $C_\ell$-fiducial model and the $C_\ell$-moments model. Prior types are either Gaussian (G) or top-hat (TH), considered distributed symmetrically around the center value with half width meaning the standard deviation (Gaussian) or the half width (top-hat). }\label{tab:priors_pipeline_a}
\begin{tabular}{c c c c c c c c c c c c c c }
\hline
\hline
\noalign{\smallskip}
model & \multicolumn{9}{c}{$C_\ell$-fiducial and $C_\ell$-moments} & \multicolumn{4}{c}{$C_\ell$-moments only}\\
\noalign{\smallskip}
\hline
\noalign{\smallskip}
parameter & $A_{\rm lens}$ & $r$ & $A_d$ & $\alpha_d$ & $\beta_d$ & $A_s$  & $\alpha_s$ & $\beta_s$ & $\epsilon_{ds}$ & $B_s$ & $\gamma_s$ & $B_d$ & $\gamma_d$\\
\noalign{\smallskip}
prior type & TH & TH & TH & TH & G & TH & TH & G & TH & TH & TH & TH & TH\\
center value & $1.0$ & $0.0$ & $25$ & $0.0$ & $1.54$ & $2.0$ & $-1.0$ & $-3.0$ & $0.0$ & $0.0$ & $-4.0$ & $5.0$ & $-4.0$\\
half width & $1.0$ & $0.1$ & $25$ & $0.5$ & $0.11$ & $2.0$ & $1.0$ & $0.3$ & $1.0$ & $10.0$ & $2.0$ & $5.0$ & $2.0$\\
\hline
\noalign{\smallskip}
\hline
\end{tabular}
\end{table*}

\subsection{Pipeline B: NILC cleaning}\label{sec:nilc}

Our second pipeline is based on the blind Internal Linear Combination (ILC) method, which assumes no information on foregrounds whatsoever, and instead only assumes that the observed data contains one signal of interest (the CMB), plus noise and contaminants \citep{2003ApJS..148...97B}. The method assumes a simple model for the observed multi-frequency maps $\mathbf{d}_{\nu}$ at $N_{\nu}$ frequency channels (in either pixel or harmonic space)
\begin{equation}
    \mathbf{d}_{\nu} = a_{\nu} \mathbf{s} + \mathbf{n}_{\nu} \text{,}
\end{equation}
where $a_{\nu}$ is the black-body spectrum of the CMB, $\mathbf{s}$ is the amplitude of the true CMB signal, and $\mathbf{n}_{\nu}$ is the contamination in channel $\nu$, which includes the foregrounds and instrumental noise. ILC exploits the difference between the black-body spectrum of the CMB and the SED(s) of other components that may be present in the data. The method aims at reconstructing a map of the CMB component $\tilde{\mathbf{s}}$ as a linear combination of the data with a set of weights $w_{\nu}$ allowed to vary across the map,
\begin{equation}\label{eq:ilc_reconstruction}
    \tilde{\mathbf{s}} = \sum_{\nu} w_{\nu} \mathbf{d}_{\nu} = \mathbf{w}^{T} \hat{\sf d}\, ,
\end{equation}
where both $\mathbf{w}$ and $\hat{\sf d}$ are $N_\nu\times N_{\rm pix}$ matrices, with $N_{\rm pix}$ being the number of pixels. We optimize the weights by minimizing the variance of $\tilde{\bf s}$ and find
\begin{equation} \label{eq:nilc_weights}
    \mathbf{w}^T = \frac{\mathbf{a}^T \hat{\sf C}^{-1} }{\mathbf{a}^T \hat{\sf C}^{-1} \mathbf{a} } \text{,} 
\end{equation}
where $\mathbf{a}$ is the black-body spectrum of the CMB (i.e., a vector filled with ones if maps are in thermodynamic temperature units), and $\hat{\sf C} = \langle \hat{\sf d}\,\hat{\sf d}^T \rangle$ is the frequency-frequency covariance matrix per pixel of the observed data. We do not assume any correlation between pixels in this work.

\begin{figure}
    \centering
    \includegraphics[width=\columnwidth]{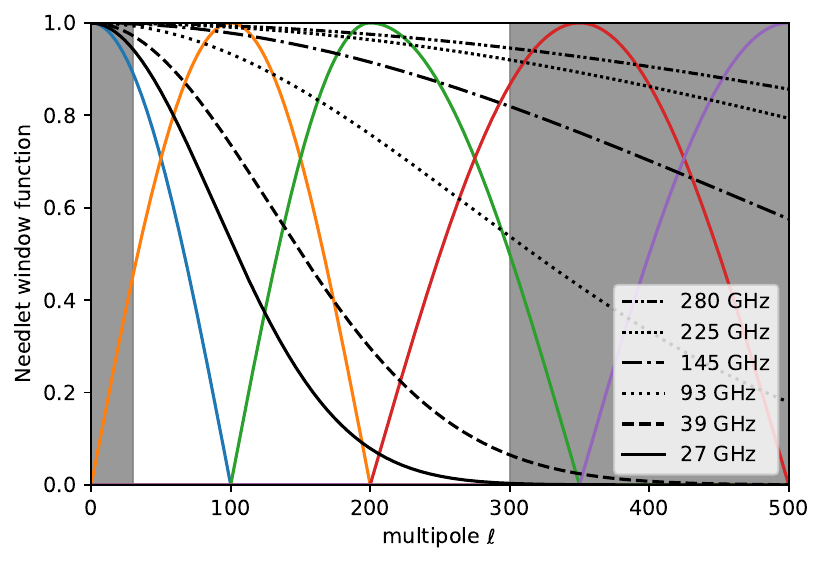}
    \caption{Five needlet windows and beam transfer functions as used in pipeline B. Colored lines are needlet windows in harmonic space, black dashed lines are the transfer functions $b_{\ell}^{\nu}$ for the six SO-SAT frequency channels. The FWHM of every beam is listed in Table~\ref{tab:SAT_instrument}. Gray shaded areas denote multipole ranges that are omitted during $r$ inference with pipeline B.} \label{fig:nilc_windows}
\end{figure}

In our particular implementation, we use the needlet internal linear combination method \citep[NILC,][]{2009A&A...493..835D,2012MNRAS.419.1163B,2013MNRAS.435...18B}. NILC uses localization in pixel and harmonic space by finding different weights ${\bf w}$ for a set of harmonic filters, called ``needlet windows''. These windows are defined in harmonic space $h_{i}(\ell)$ for $i=0,...,n_{\rm windows}-1$ and must satisfy the constraint $\sum_{i=0}^{n_{\rm windows}-1} h_{i}(\ell)^2 = 1$ in order to preserve the power of the reconstructed CMB. We use $n_{\rm windows}=5$ needlet windows shown in Fig.~\ref{fig:nilc_windows}, and defined by
\small
\begin{equation}
    h_{i}(\ell) = 
    \begin{cases}
        \cos(\frac{\pi}{2}(\ell^{\rm peak}_i - \ell)/(\ell^{\rm peak}_i - \ell^{\rm min}_i)) & \text{if $\ell^{\rm min}_i \leq \ell < \ell^{\rm peak}_i$ }\\
        1 & \text{if $\ell = \ell^{\rm peak}_i$} \\
        \cos(\frac{\pi}{2}(\ell-\ell^{\rm peak}_i)/(\ell^{\rm max}_i-\ell^{\rm peak}_i)) & \text{if $\ell^{\rm peak}_i < \ell \leq \ell^{\rm max}_i$}
    \end{cases}
    \text{,}
\end{equation}
\normalsize
with $\ell_{\rm min}=\{$0, 0, 100, 200, 350$\}$, $\ell_{\rm max}=\{$100, 200, 350, 500, 500$\}$, and $\ell_{\rm peak}=\{$0, 100, 200, 350, 500$\}$ for the corresponding five needlet windows. Even though we do not use the full 500-$\ell$ range where windows are defined for the likelihood sampling, we still perform the component separation on all five windows up to multipoles beyond our upper limit of $\ell=300$, in order to avoid edge effects on the smaller scales.

\begin{figure}
\centering
 \includegraphics[width=\columnwidth]{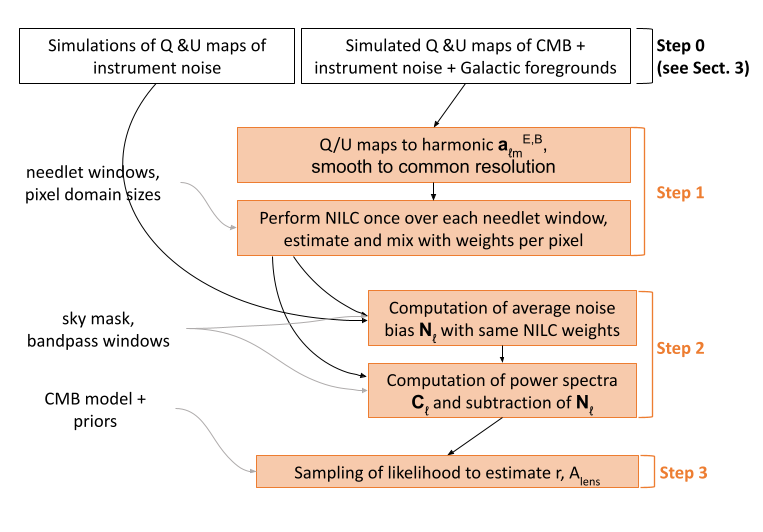}
\caption{Schematic of pipeline B. Orange colors mark steps that are repeated 500 times, once for each simulation.} \label{fig:schematic_b}
\end{figure}

Let us now describe the NILC procedure as illustrated in Fig.~\ref{fig:schematic_b}. In step 1, we perform our CMB reconstruction in the $E$ and $B$ field instead of $Q$ and $U$. We transform the observed maps to $a_{\ell m}^X$ with $X \in E,B$. All frequency channels are then brought to a common beam resolution by rescaling the harmonic coefficients with an appropriate harmonic beam window function. The common beam we adopt is the one from the third frequency channel at 93\,GHz, which corresponds to a FWHM of 30\,arcmin. For each needlet window index $i$, we multiply $a_{\ell m}^X$ by $h_i(\ell)$ as a harmonic filter. Since different frequency channels have different limiting resolutions, we do not use all channels in every needlet window. The first two windows use all six frequency channels, the third window does not use the 27 GHz channel, and the last two needlet windows do not use the 27 and 39 GHz channels. The covariance matrix $\hat{\sf C}$ has dimensions $N_\nu\times N_\nu\times N_{\rm pix}$. For each pixel $p$, its corresponding $N_\nu\times N_\nu$ elements are computed directly from the data, averaging over the pixels inside a given pixel domain $\mathcal{D}(p,i)$ around each pixel. In practice, the element $\nu,\nu'$ of the covariance matrix is calculated by multiplying the two filtered maps at channels $\nu$ and $\nu'$, then smoothing that map with a Gaussian kernel with FWHM equal to the size of the pixel domain $\mathcal{D}(p,i)$. The FWHMs for the pixel domain size are 185, 72, 44, 31, and 39 degrees for each needlet window, respectively. \footnote{The domain sizes are estimated directly from the needlet window scale \citep[see details in the appendix A of ][]{2009A&A...493..835D}. The ILC bias can be minimized by enlarging the pixel domains to be big enough to include a higher number of modes. We choose the resulting ILC bias to not exceed $0.2\%$, for which we need pixel domain sizes large enough so that each needlet window contains at least 2500 modes.}

We then proceed to calculate the weights $\mathbf{w}^T$ (see Eq.~\ref{eq:nilc_weights}) for window $i$, which is an array with shape (2, $N_{\nu}$, $N_{\rm pixels}^{i}$), with the first dimension corresponding to the $E$ and $B$ fields. We note that the number of pixels $N_{\rm pixels}^i$ is different for each needlet window, since we use different pixel resolutions that depend on the smallest scale covered by the respective window. Finally, we apply Eq.~\ref{eq:ilc_reconstruction} to obtain an ILC-reconstructed CMB map for window $i$. The final step is to filter this map in harmonic space for a second time with the $h_i(\ell)$ window. The final reconstructed CMB map is the sum of these maps for all five needlet windows.

In step 2, the reconstructed CMB maps are compressed into power spectra using \texttt{NaMaster} and deconvolved to the common beam resolution. We use B-mode purification as implemented in the software and the mask shown in Fig.~\ref{fig:mask}. We estimate the noise bias $N_{\ell}$ in the final map by computing the power spectrum of noise-only simulations processed with the needlet weights and windows obtained from the simulated data as described above. $N_{\ell}$ is averaged over simulations and subtracted from the $C_{\ell}$ of the reconstructed maps. 

Finally, in step 3, we run a Monte Carlo Markov Chain (MCMC) over the reconstructed $BB$ spectrum (we ignore $EE$ and $EB$) with two free parameters, the tensor-to-scalar ratio $r$ and the amplitude of the $BB$ lensing spectrum $A_{\rm lens}$. For the posterior sampling, we use the Python package \texttt{emcee}. Both parameters have a top hat prior (between 0 and 2 for $A_{\rm lens}$, and between $-0.013$ and infinity for $r$). The covariance matrix is calculated directly over 500 simulations with the same setup but with Gaussian foregrounds. As likelihood, we use the same Gaussian likelihood used in pipeline A and restrict the inference to a multipole range $30 < \ell \leq 300$.

While the NILC implementation described above is blind, it can be extended to a semi-blind approach that introduces a certain level of foreground modeling. For example, constrained ILC \citep[cILC,][]{2011MNRAS.410.2481R} explicitly nullifies one or more contaminants (such as thermal dust) in observed maps, by including their modeled SED in the variance minimization that results in the ILC weights (see Eq.~\ref{eq:nilc_weights}). This foreground modeling can be further extended to include the moment expansion of the SED described in Sect.~\ref{ssec:pipelines.Cell}. This method, known as constrained moment ILC \citep[cMILC][]{2021MNRAS.503.2478R}, has proven effective at cleaning the large-scale B-mode contamination for space experiments such as LiteBIRD. While not used in this work, these extensions and others will be considered in future analyses with more complex foregrounds and systematics.

\subsection{Pipeline C: map-based cleaning}

\begin{figure}
\centering
 \includegraphics[width=\columnwidth]{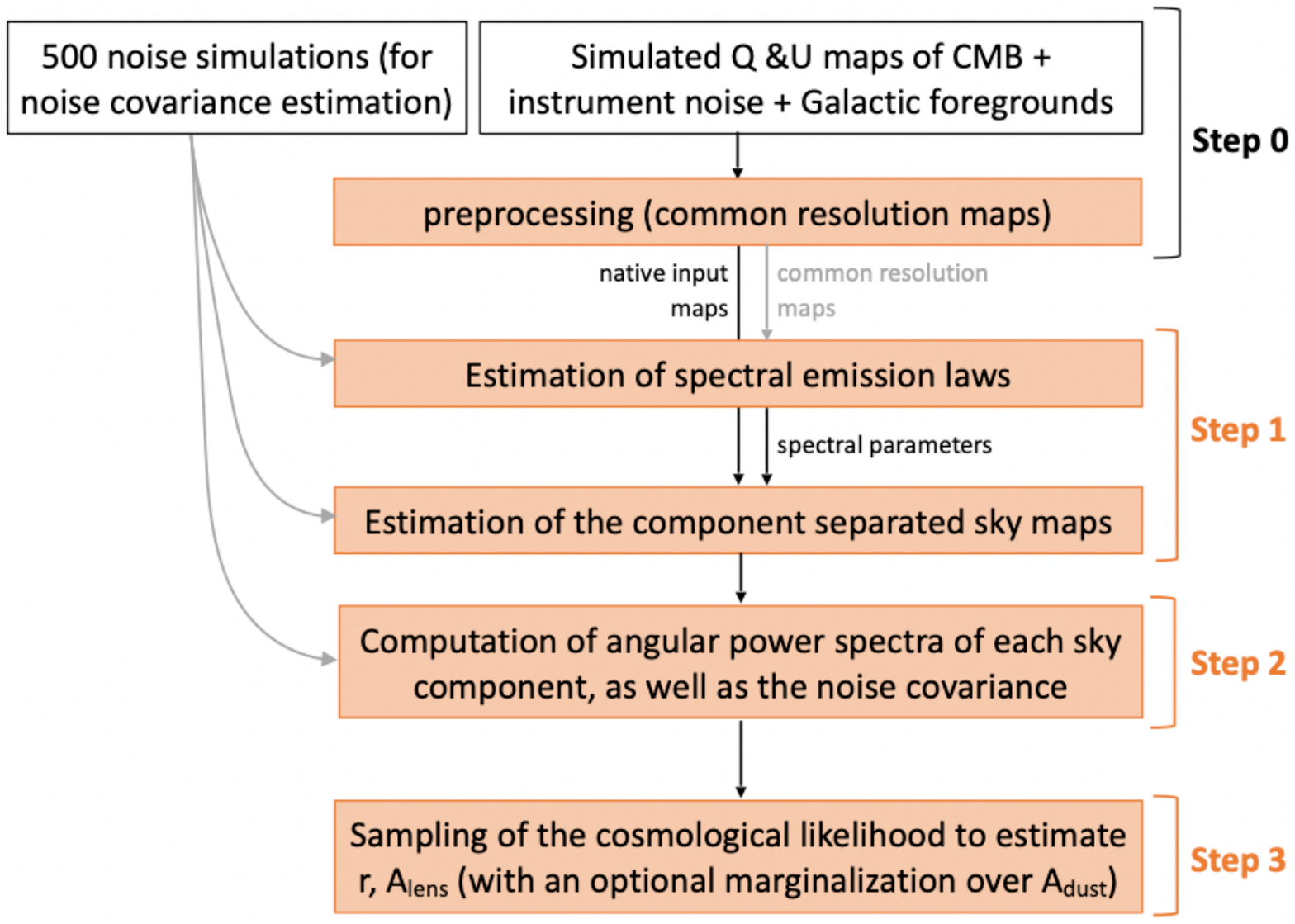}
\caption{Schematic of pipeline C. Orange colors indicate repetition for each simulation.} \label{fig:schematic_c}
\end{figure}

Our third pipeline is a map-based parametric pipeline based on the \texttt{fgbuster} code~\citep{Poletti_Errard_2022}. This approach is based on the data model 
\begin{eqnarray}
    \centering 
        {\bf d} = \hat{\sf A}{\bf s} + {\bf n} \, ,
\end{eqnarray}
where ${\bf d}$ is a vector containing the polarized frequency maps, ${\bf s}$ is a vector containing the $Q$ and $U$ amplitudes of the sky signals (CMB, foregrounds) and ${\bf n}$ is the noise contained in each frequency map. The matrix $\hat{\sf A}=\hat{\sf A}(\boldsymbol{\beta})$ is the so-called mixing matrix, assumed to be parameterized by a set of spectral indices $\boldsymbol{\beta}$. Starting from the observed (mock) input data ${\bf d}$, Fig.~\ref{fig:schematic_c} shows a schematic of the pipeline, comprising four steps. 

Step 0 is the preprocessing of input simulations. For each simulation, we combine the simulated noise maps, the foreground, and CMB maps and save them on disk. We create a new set of frequency maps, $\mathbf{\bar{d}}$, smoothed with a common Gaussian kernel of $100\arcmin$ FWHM.

Step 1 is the actual component separation stage. We optimize the spectral likelihood, defined as~\citep{Stompor2009}:
\begin{align}\label{eq:spectral_likelihood}
  & -2\log\left(\mathcal{L}_{\rm spec}(\boldsymbol{\beta})\right) = \notag \\
  & \qquad \left(\hat{\sf A}^T\hat{\sf N}^{-1}\bar{\bf d}\right)^T\left(\hat{\sf A}^T\hat{\sf N}^{-1}\hat{\sf A}\right)^{-1}\left(\hat{\sf A}^T\hat{\sf N}^{-1}\bar{\bf d}\right)
\end{align}
which uses the common resolution frequency maps, $\bar{\bf d}$, built during step 0. The right hand side of Eq.~\ref{eq:spectral_likelihood} contains a sum over the observed sky pixels, assumed to have uncorrelated noise. The diagonal noise covariance matrix $\hat{\sf N}$ is computed from 500 noise-only simulations. Although, in principle, $\hat{\sf N}$ can be non-diagonal, we do not observe any significant bias of the spectral likelihood due to this approximation in this study. By minimizing Eq.~\ref{eq:spectral_likelihood} we estimate the best-fit spectral indices $\tilde{\boldsymbol{\beta}}$ and the corresponding mixing matrix $\tilde{\sf A} \equiv \hat{\sf A}(\tilde{\boldsymbol{\beta}})$. We also estimate the uncertainties on the recovered spectral indices as provided by the minimizer, a truncated Newton algorithm~\citep{1984SJNA...21..770N} as implemented in \texttt{scipy}~\citep{2020NatMe..17..261V}.
Having thus obtained estimators of the foreground SEDs, we can recover the sky component maps with the generalized least-square equation
\begin{equation}\label{eq:weights_c}
   \tilde{\bf s} = \left(\tilde{\sf A}^T\hat{\sf N}^{-1}\tilde{\sf A}\right)^{-1}\tilde{\sf A}^T\hat{\sf N}^{-1}{\bf d} \equiv \hat{\sf W}{\bf d}\; ,
\end{equation}
where ${\bf d}$ is the input raw data, and not the common resolution maps.
In steps 1 and 2, we have the possibility to use an inhomogeneous noise covariance matrix $\hat{\sf N} = \hat{\sf N}(\nv)$ and, although this is not exploited in this work, a spatially varying mixing matrix $\boldsymbol{\beta} = \boldsymbol{\beta}(\nv)$. For the latter, one can use the multi-patch or clustering methods implemented in \texttt{fgbuster}~\citep{Errard2019,Puglisi2022}.

Step 2 comprises the calculation of angular power spectra. The recovered CMB polarization map is transformed to harmonic space using \texttt{NaMaster}. We estimate an effective transfer function, ${\bf B}^{\rm eff}_\ell=\hat{\sf W}{\bf B}_\ell$, associated with the reconstructed components $\tilde{\bf s}$, from the channel-specific beams ${\bf B}_\ell$. Correcting for the impact of this effective beam is vital to obtain an unbiased $BB$ power spectrum of the foreground-cleaned CMB, $\tilde{C}_\ell^{\rm CMB}$. In the second step, we use noise simulations to estimate the noise bias
\begin{equation}
  \tilde{\sf N}_\ell = \frac{1}{\rm N_{sim}}\sum_{\rm sims}\sum_{m=-\ell}^{\ell}\frac{\tilde{\bf n}_{\ell,m} \tilde{\bf n}_{\ell,m'}^\dagger}{2\ell+1}.
\end{equation}
where $\tilde{\bf n} = \hat{\sf W}{\bf n}^{\rm sim}$ is the noise in the recovered component-separated sky maps. We consider 500 simulations to estimate the noise bias.

Step 3 is the cosmological analysis stage. We model the angular power spectrum of the component-separated CMB map, including the noise contribution, as
\begin{equation}
    C_\ell^{\rm CMB}(r,A_{\rm lens}) \equiv C_\ell^{\rm prim}(r)+C_\ell^{\rm lens}(A_{\rm lens}) + \tilde{N}^{\rm CMB}_\ell
    \label{eq:cosmo_model}
\end{equation}
and compare data and model with the cosmological likelihood
\begin{align}
        & -2\log{\mathcal{L}^{\rm cosmo}} \notag \\ 
        & \qquad =\sum_\ell \left(2\ell + 1 \right)f_{\rm sky}\left( \frac{\tilde{C}_\ell^{\rm CMB}}{C_\ell^{\rm CMB}} + \log(C_\ell^{\rm CMB})\right).
    \label{eq:cosmo_likelihood}
\end{align}
It is worth noting that this is only an approximation to the true map-level Gaussian likelihood which approximates the effective number of modes in each multipole after masking and purification as $f_{\rm sky}(2\ell+1)$, thus neglecting any mode-coupling effects induced by the survey footprint.
We grid the likelihood above along the two dimensions $r$ and $A_{\rm lens}$. For each simulation we then estimate the maximum-likelihood values and 68\% credible intervals from the marginal distributions of $r$ and $A_{\rm lens}$. We verified that the distributions of recovered $\{r,\,A_{\rm lens}\}$ across simulations are well described by a Gaussian, hence supporting the Gaussian likelihood in Eq.~\ref{eq:cosmo_likelihood}. \\

Pipeline C also offers the option to marginalize over a dust template. The recovered components in $\tilde{\bf s}$, Eq.~\ref{eq:weights_c}, include the dust $Q$ and $U$ maps which are typically recovered with high signal-to-noise. In the same way that we compute $\tilde{C}_\ell^{\rm CMB}$ in step 2, we compute the $BB$ component of the recovered dust map, $\tilde{C}_\ell^{\rm dust}$.
We then update our cosmological likelihood, Eq.~\ref{eq:cosmo_model}, by adding a dust term:
\begin{equation}
  C_\ell^{\rm CMB} = C_\ell^{\rm CMB}(r,\,A_{\rm lens}) + A_{\rm dust}\tilde{C}_\ell^{\rm dust}.\label{eq:cosmo_model_w_dust}
\end{equation}
This is a similar approach to earlier methods~\citep{Errard2019,PTEP2022}. When choosing this approach, the inference of $r$ during step 3 therefore involves the marginalization over both parameters $A_{\rm lens}$ and $A_{\rm dust}$. In principle one could add synchrotron or other terms in Eq.~\ref{eq:cosmo_model_w_dust} but we limit ourselves to dust as it turns out to be the largest contamination, and, in practice, marginalizing over it allows us to get unbiased estimates of cosmological parameters. In the remainder of this paper, we refer to this method as ``C + dust marginalization''.

\section{Description of input simulations}\label{sec:sims}

\begin{table*}
  \caption{Instrument and noise specifications used to produce the simulations in this work. It should be stressed that these levels correspond to homogeneous noise, while our default analysis assumes noise maps weighted according to the SAT hits map.}\label{tab:SAT_instrument}
   \begin{threeparttable}
     \begin{tabular}{ccccccc} 
        \hline\hline
        \noalign{\smallskip}\noalign{\smallskip}
        && Baseline & Goal  & Pessimistic & Optimistic & \\
        \noalign{\smallskip}\noalign{\smallskip}
    Frequency [GHz] & FWHM [arcmin] & Noise [$\mu$K-arcmin] & Noise [$\mu$K-arcmin]  & $\ell_{\rm knee}$ & $\ell_{\rm knee}$ & $\alpha_{\rm knee}$ \\
    
        \noalign{\smallskip}
         \hline
        \noalign{\smallskip}
          27 & 91 & 46 & 33 & 30 & 15 & -2.4\\
          39 & 63 & 28 & 22 & 30 & 15 & -2.4\\
          93 & 30 & 3.5 & 2.5 & 50 & 25 & -2.5\\
          145 & 17 & 4.4 & 2.8 & 50 & 25 & -3.0 \\
          225 & 11 & 8.4 & 5.5 & 70 & 35 & -3.0 \\
          280 & 9 & 21 & 14 & 100 & 40 & -3.0 \\
        \hline
         \noalign{\smallskip}
        \hline
     \end{tabular}
 \end{threeparttable}
\end{table*}
We built a set of dedicated simulations against which to test our data analysis pipelines and compare results. The simulated maps include cosmological CMB signal, Galactic foreground emission as well as instrumental noise.

\subsection{Instrumental specifications and noise}\label{ssec:sims.inst}

We simulate polarized Stokes $Q$ and $U$ sky maps as observed by the SO-SAT telescopes. All maps are simulated using the \texttt{HEALPix} pixelation scheme \citep{2005ApJ...622..759G} with resolution parameter $N_{\rm side}=512$.

We model the SO-SAT noise power spectra as
\begin{equation}
  N_{\ell} = N_{\rm white}\Big[1+\Big(\frac{\ell}{\ell_{\rm knee}}\Big)^{\alpha_{\rm knee}}\Big],
\end{equation}
where $N_{\rm white}$ is the white noise component while $\ell_{\rm knee}$, and $\alpha_{\rm knee}$ describe the contribution from $1/f$ noise. Following \cite{2019JCAP...02..056A} (hereinafter SO2019), we consider four scenarios: ``baseline'' and ``goal'' levels for the white noise component, and ``pessimistic'' and ``optimistic'' correlated noise. The empirical $1/f$ scenarios are based on measurements from recent experiments and consider polarization modulation, filtering, and atmospheric transmission corresponding to the conditions at the Atacama site. The values of white noise, $\ell_{\rm knee}$, and $\alpha_{\rm knee}$ associated with the different cases are reported in Table~\ref{tab:SAT_instrument}. We note that noise levels correspond to a sky fraction of $f_{\rm sky}=10\%$ and five years of observation time, as in SO2019. Differently from SO2019, we cite polarization noise levels at a uniform map coverage, accounting for the factor of $\sim1.3$ difference compared to Table~1 in SO2019.\footnote{Polarization noise accounts for a factor of $\sqrt{2}$ and homogeneous noise for a factor of $\sqrt{0.85}$ compared to Table~1 in SO2019.} We simulate noise maps as Gaussian realizations of the $N_{\ell}$ power spectra. In our main analysis, we use noise maps with pixel weights computed from the SO-SAT hits map (see Fig.~\ref{fig:mask}) and refer to this as ``inhomogeneous noise''. In Sect.~\ref{ssec:results.r}, we briefly present results obtained from equally weighted noise pixels, which we refer to as ``homogeneous noise''. Otherwise, all results in this paper assume inhomogeneous noise. We note that, although inhomogeneous, the noise realizations used here lack some of the important anisotropic properties of realistic $1/f$ noise, such as stripes due to the scanning strategy. Thus, together with the impact of other time-domain effects (e.g. filtering), we leave a more thorough study of the impact of instrumental noise properties for future work. 

\subsection{CMB}\label{ssec:sims.cmb}

We simulate the CMB signal as isotropic Gaussian random realizations following a power spectrum computed at the \textit{Planck} 2018 best-fit $\Lambda$CDM cosmology. Our baseline model does not include any primordial tensor signal ($r=0$) but incorporates lensing power in the $BB$ spectra ($A_{\rm lens}=1$). We consider also two modifications of this model: (i) primordial tensor signal with $r=0.01$, representing a $\gtrsim 3\sigma$ target detection for SO with $\sigma(r)=0.003$, as forecasted by SO2019; (ii) reduced lensing power with $A_{\rm lens}=0.5$, corresponding to a 50\% delensing efficiency, achievable for SO as shown in \cite{2022PhRvD.105b3511N}.

For every scenario, we simulated 500 realizations of the CMB signal, convolved with Gaussian beams for each frequency channel, with FWHMs as reported in Table~\ref{tab:SAT_instrument}. 

\subsection{Foregrounds}\label{ssec:sims.fg}

Thermal emission from Galactic dust grains and synchrotron radiation are known to be the two main contaminants to CMB observations in polarization, at intermediate and large angular scales, impacting therefore measurements of the primordial $BB$ signal. The past years have seen many studies on the characterization of polarized Galactic foreground emission, thanks to the analysis of \textit{WMAP} and \textit{Planck} data, as well as low frequency surveys \citep{2022MNRAS.513.5900H, 2018A&A...618A.166K}. However, many aspects of their emission remain unconstrained, including, in particular, the characterization of their SEDs and their corresponding variation across the sky. To properly assess the impact of foreground emission on component separation and $r$ constraints, we therefore use four sets of sky emission models. As specified in the following, we use the Python sky model ({\sc PySM}) package \citep{2017MNRAS.469.2821T} to simulate polarized foreground components, with some additional modifications: 

\begin{itemize}
    \item {\bf Gaussian foregrounds}: we simulate thermal dust emission and synchrotron radiation as Gaussian realizations of power law $EE$ and $BB$ power spectra. Although inaccurate, since foregrounds are highly non-Gaussian, this idealistic model was used to validate the different pipelines and to build approximate signal covariance matrices from 500 random realizations. In particular, we estimate the amplitudes of the polarized foreground signal (evaluated for $D_{\ell} = \ell(\ell+1)C_{\ell}/2\pi$ at $\ell=80$) and the slope of angular power spectra from the {\sc PySM} synchrotron and thermal dust templates, evaluated at the SO-SAT sky patch. We obtain the following values ($d$: thermal dust at 353 GHz; $s$: synchrotron at 23 GHz): $A^d_{EE} = 56$ $\mu K_{\rm CMB}^{2}$, $A^d_{BB} = 28$ $\mu K_{\rm CMB}^2$, $\alpha^d_{EE} = -0.32$, $\alpha^d_{BB} = -0.16$; $A^s_{EE} = 9$ $\mu K_{\rm CMB}^{2}$, $A^s_{BB} = 1.6$ $\mu K_{\rm CMB}^2$, $\alpha^s_{EE} = -0.7$, $\alpha^s_{BB} = -0.93$. This model assumes the frequency scaling of the maps across the SO channels to be a modified black body for thermal dust emission, with fixed spectral parameters $\beta_d=1.54$ and $T_d = 20$ K, and a power law for synchrotron with fixed $\beta_s=-3$ (in antenna temperature units).

    \item {\bf \dzsz{} model}: in this case, multi-frequency maps are taken from the \dzsz{} {\sc PySM} model. This model includes templates for thermal dust emission coming from \textit{Planck} high frequency observations and from \textit{WMAP} 23 GHz maps from synchrotron radiation. SEDs are considered to be uniform across the sky with the same values of the spectral parameters used for the Gaussian simulations.

    \item{\bf \doso{} model}: this model uses the same foreground amplitude templates as \dzsz{}, but with the inclusion of spatial variability for spectral parameters, as described in \cite{2017MNRAS.469.2821T}.

    \item{\bf \dmsm{} model}: this model represents a modification of the \doso{} spatial variation of spectral parameters. For thermal dust we smoothed the $\beta_d$ and $T_d$ templates at an angular resolution of 2 degrees, in order to down-weight the contribution of instrumental noise fluctuations in the original {\sc PySM} maps. For synchrotron emission we modified the $\beta_s$ {\sc PySM} in order to account for the additional information coming from the analysis of S-PASS data at 2.3 GHz (see \cite{2018A&A...618A.166K}). In particular S-PASS data show that the synchrotron spectral index presents enhanced variations with respect to the {\sc PySM} template. We therefore multiplied the fluctuations in the $\beta_s$ map by a factor 1.6 to take into consideration larger variations. Moreover, we added small scale fluctuations (with a minimum angular resolution of 2 degrees), as Gaussian realization of a power-law power spectrum with slope $-2.6$ (see Fig.~11 in \cite{2018A&A...618A.166K}).
\end{itemize}

We note that this set of foreground models generalizes the what was done \cite{2019JCAP...02..056A}, since it includes the \doso{} model, used for large-scale B-mode forecasts in that earlier analysis. As for the CMB simulations, the multi-frequency foreground maps at the SO reference frequencies were convolved with Gaussian beams, to reach the expected angular resolution. We assumed delta-like frequency bandpasses in order to accelerate the production of these simulations, although all pipelines are able to handle finite bandpasses. Therefore this approximation should not impact the performance of any of the pipelines presented here.

\section{Results and discussion}\label{sec:results}

Simulations were generated for four different noise and foreground models, respectively, (see Sect.~\ref{sec:sims}), for a total of 16 different foreground-noise combinations. For the main analysis, we consider a fiducial CMB model with $A_{\rm lens}=1$ (no delensing) and $r=0$ (no primordial tensor fluctuations). In addition, we explored three departures from the fiducial CMB model, with input parameters $(A_{\rm lens}=0.5,\,r=0)$, $(A_{\rm lens}=1,\,r=0.01)$, and $(A_{\rm lens}=0.5,\,r=0.01)$. Here we report the results found for all these cases.

\subsection{Power spectra}\label{ssec:results.cls}

Let us start by examining the CMB power spectrum products. Pipelines B and C produce CMB-only maps and base their inference of $r$ on the resulting power spectra, whereas pipeline A works directly with the cross-frequency power spectra of the original multi-frequency maps. Nevertheless, CMB power spectra are an important data product that every pipeline should be able to provide. Following the methods presented in \cite{Dunkley:2013vu} and \cite{planck_xi_2015,planck_v_2020}, we use a modified version of pipeline A that retrieves CMB-only bandpowers from multi-frequency power spectra, marginalizing over foregrounds with an MCMC sampler as presented in Sect.~\ref{ssec:pipelines.Cell}. We note that this method, originally developed for high-$\ell$ CMB science, is applicable since we are in the Gaussian likelihood regime. By re-inserting this cleaned CMB spectrum into a Gaussian likelihood with parameters $(r,\,A_{\rm lens})$, we obtain constraints that are consistent with the results shown in Table~\ref{tab:r_fiducial_results}.

\begin{figure}
    \centering
    \includegraphics[width=\columnwidth]{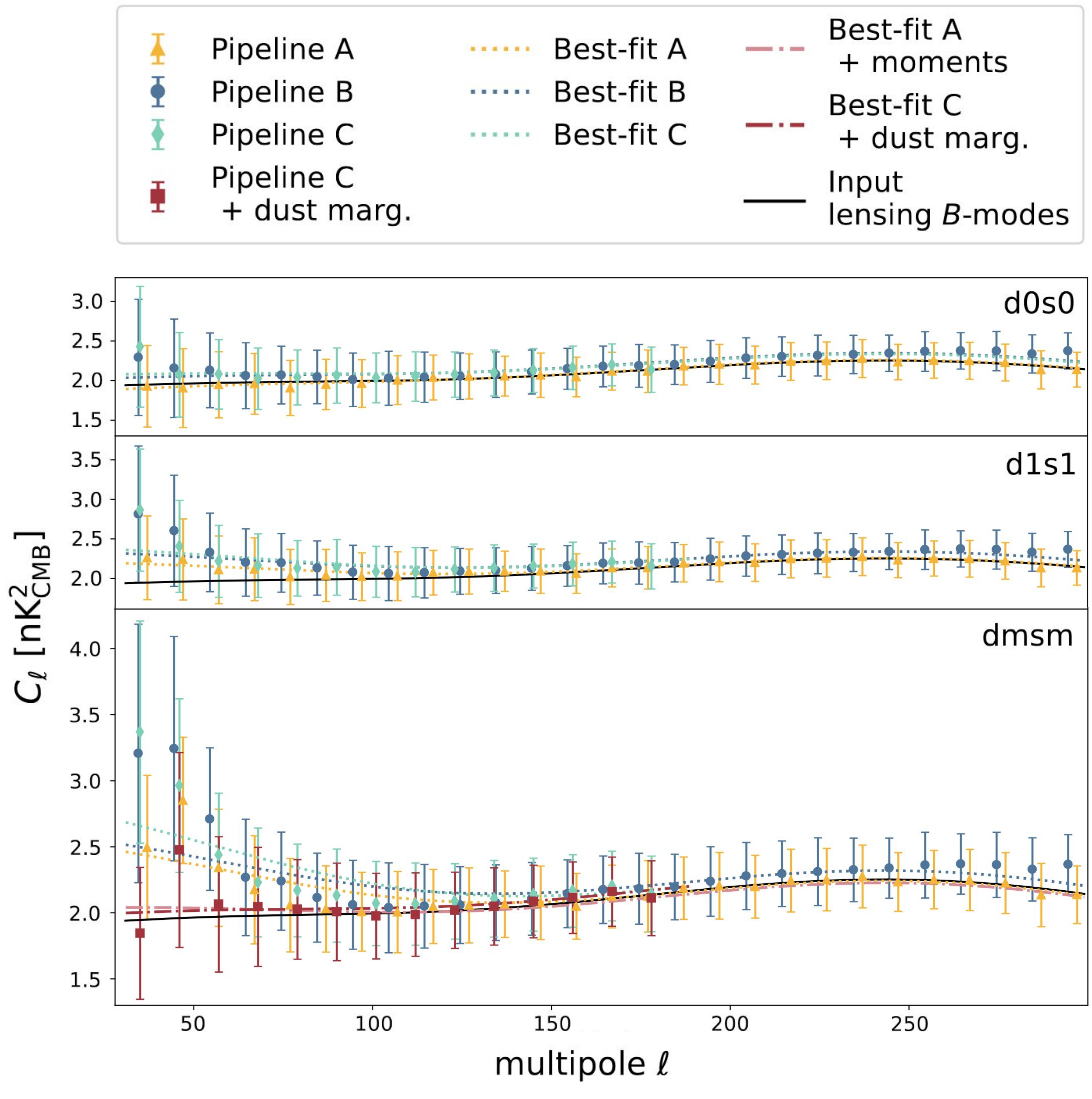}
    \caption{CMB-only power spectra resulting from component separation with pipelines A, B, and C. We show non-Gaussian foregrounds scenarios \dzsz{} (\emph{top panel}), \doso{} (\emph{middle panel}), and \dmsm{} (\emph{bottom panel}) and consider the goal-optimistic noise scenario. The different colored markers with error bars show the mean of 500 simulations and the scatter between them (corresponding to the statistical uncertainties of a single realization). The dotted lines in the corresponding colors indicate the best-fit power spectrum model. In the \dmsm{} case, we show the extended pipeline results from A + moments and C + dust marginalization with the best-fit models shown as dot-dashed lines. The black solid line is the input CMB model containing lensing B-modes only. We stress that pipeline C only considers multipoles up to $\ell=180$ in the power spectrum likelihood.}\label{fig:cells-cmb}
\end{figure}

Figure~\ref{fig:cells-cmb} shows the CMB power spectra for the three complex foreground simulations \dzsz{}, \doso{}, and \dmsm{} (upper, middle, and lower panel, respectively) while considering the goal-optimistic noise scenario. The various markers with error bars denote the measured CMB power spectra and their $1\sigma$ standard deviation across 500 simulations, while the black solid line denotes the input CMB power spectrum. Results are shown in gold triangles, blue circles, turquoise diamonds for pipeline A, B, and C respectively. The dotted lines show the best-fit CMB model for the three nominal pipelines (using the same color scheme). Only in the \dmsm{} foreground scenario, which is the most complex considered here, we also show the results from pipeline C + dust marginalization (dark red squares with error bars), and the best-fit CMB power spectrum from A + moments (pink dot-dashed line) and C + dust marginalization (dark red dot-dashed line).
    
For the nominal pipelines (A, B, and C) without extensions, the measured power spectra display a deviation from the input CMB at low multipoles, increasing with rising foreground complexity. For \dmsm{} at multipoles $\lesssim 50$, this bias amounts to about 1.5$\sigma$ and goes down to less than 0.5$\sigma$ at $80\lesssim \ell\lesssim 250$. The three pipelines agree reasonably well, while pipeline A appears slightly less biased for the lowest multipoles. Pipelines B and C show an additional mild excess of power in their highest multipole bins, with a $<0.3\sigma$ increase in pipeline C for $130 \lesssim \ell\lesssim 170$ and up to 1$\sigma$ for the highest multipole ($\ell=297$) in pipeline B. This might indicate power leakage from the multiple operations on map resolutions implemented in pipelines B and C. In pipeline B, these systematics could come from first deconvolving the multi-frequency maps and then convolving them with a common beam in order to bring them to a common resolution, whereas in pipeline C, the leakage is likely due to the linear combination of the multi-resolution frequency maps following Eq.~\eqref{eq:weights_c}. Other multipole powers lie within the 1$\sigma$ standard deviation from simulations for all three pipelines.
    
Both extensions, A + moments and C + dust marginalization, lead to an unbiased CMB power spectrum model, as shown by the pink and dark red dot-dashed lines and the square markers in the lower panel of Fig.~\ref{fig:cells-cmb}. In the case of pipelines B and C, comparing the best-fit models obtained from the measured power spectra to the input CMB model, we find sub-sigma bias for all bins with $\ell>100$. We show, however, that the ability to marginalize over additional foreground residuals (e.g. the dust-template marginalization in pipeline C) is able to reduce this bias on all scales, at the cost of increased uncertainties. Implementing this capability in the blind NILC pipeline B would likely allow to reduce the bias that we see.

The SO-SATs are expected to constrain the amplitude of CMB lensing B-modes to an unprecedented precision. As can be seen from Fig.~\ref{fig:cells-cmb}, individual, cleaned CMB bandpowers without delensing at multipoles $\ell\gtrsim150$ achieve a signal-to-noise ratio of about 10, accounting for a combined precision on the lensing amplitude of $\sigma(A_{\rm lens})\lesssim0.03$ when considering multipoles up to $\ell_{\rm max}=300$. As we show in the following section, this is consistent with the inference results obtained by pipelines A and B.

\subsection{Constraints on $r$}\label{ssec:results.r}

\begin{figure}
    \centering
    \includegraphics[width=\columnwidth]{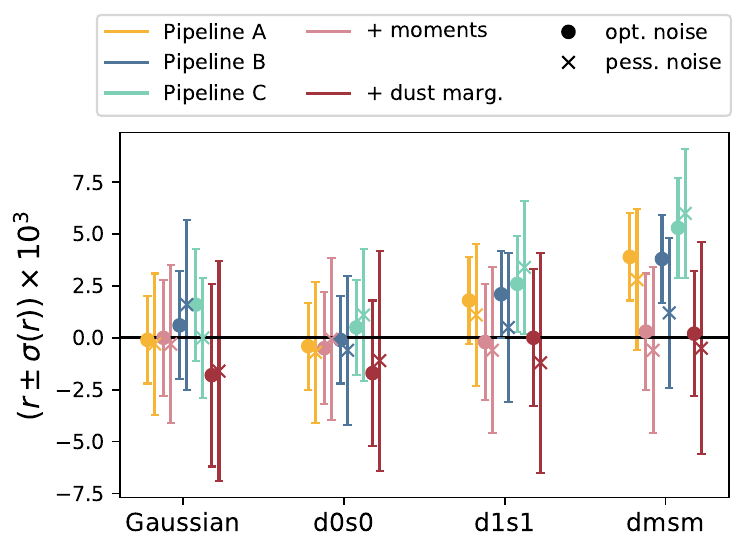}
    \caption{Mean $r$ with $(16,\,84)$\% credible interval from 500 simulations. We apply the three nominal component separation pipelines (plus extensions) to simulations with four foreground scenarios of increasing complexity. We assume a fiducial cosmology with $r=0$ and $A_{\rm lens}=1$, inhomogeneous noise with goal sensitivity and optimistic $1/f$ noise component (dot markers), and inhomogeneous noise with baseline sensitivity and pessimistic $1/f$ noise component (cross markers). We note that the NILC results for Gaussian foregrounds are based on a smaller sky mask, see Appendix~\ref{sec:nilc_gaussian_bias}.}
    \label{fig:r_sigma_r_compilation}
\end{figure}
    
\begin{table*}
\caption{Mean $r$ and $(16,\,84)$\% credible interval from 500 simulations, as inferred by three pipelines (and two extensions) on four foreground models and four noise cases, two of which are shown in Fig.~\ref{fig:r_sigma_r_compilation}. No delensing is assumed in these fiducial results.} \label{tab:r_fiducial_results}
    \centering
    \begin{tabular}{ccccccc}
    \hline
    \hline
    \noalign{\smallskip}
    \multicolumn{7}{c}{$10^3\times(r\pm\sigma(r))$}\\
    \noalign{\smallskip}
    \hline
    \noalign{\smallskip}
    Noise & FG model & Pipeline A & $+$ moments & Pipeline B & Pipeline C & $+$ dust marg.\\
    \noalign{\smallskip}
    \hline
    \noalign{\smallskip}
     & Gaussian &  $-0.1 \pm 2.1$ &  $0.0 \pm 2.8$ &  $0.6\pm2.6^{\dagger}$ &  $1.6 \pm 2.7$ &  $-1.8 \pm 4.4$ \\
    Goal rms, & \dzsz{} &  $-0.4 \pm 2.1$ &  $-0.5 \pm 2.7$ &  $-0.1 \pm 2.1$ &  $0.5 \pm 2.3$ &  $-1.7 \pm 3.5$ \\
    optimistic $1/f$ & \doso{} &  $1.8 \pm 2.1$ &  $-0.2 \pm 2.8$ &  $2.1 \pm 2.1$ &  $2.6 \pm 2.3$ &  $0.0 \pm 3.3$ \\
     & \dmsm{} &  $3.9 \pm 2.1$ &  $0.3 \pm 2.7$ &  $3.8 \pm 2.1$ &  $5.3 \pm 2.4$ &  $0.2 \pm 3.0$ \\
    \noalign{\smallskip}
    \hline
    \noalign{\smallskip}
     & Gaussian &  $-0.2 \pm 2.5$ &  $-0.1 \pm 2.7$ &  $1.1\pm3.3^{\dagger}$ &  $0.9 \pm 2.1$ &  $-0.9 \pm 5.3$ \\
    Goal rms, & \dzsz{} &  $-0.6 \pm 2.5$ &  $-0.5 \pm 2.8$ &  $-0.5 \pm 2.8$ &  $0.1 \pm 2.5$ &  $-0.9 \pm 4.0$ \\
    pessimistic $1/f$ & \doso{} &  $1.3 \pm 2.5$ &  $0.1 \pm 3.0$ &  $1.2 \pm 2.8$ &  $3.4 \pm 3.1$ &  $-0.0 \pm 3.9$ \\
     & \dmsm{} &  $3.2 \pm 2.6$ &  $0.3 \pm 3.9$ &  $2.1 \pm 2.8$ &  $5.5 \pm 2.4$ &  $0.6 \pm 4.2$ \\
    \noalign{\smallskip}
    \hline
    \noalign{\smallskip}
     & Gaussian &  $-0.1 \pm 2.6$ &  $-0.3 \pm 3.3$ &  $0.5\pm3.3^{\dagger}$ &  $0.5 \pm 3.2$ &  $-1.9 \pm 5.9$ \\
    Baseline rms, & \dzsz{} &  $-0.4 \pm 2.6$ &  $-0.3 \pm 3.3$ &  $-0.9 \pm 2.7$ &  $0.7 \pm 2.9$ &  $-1.8 \pm 4.4$ \\
    optimistic $1/f$ & \doso{} &  $1.7 \pm 2.6$ &  $-0.2 \pm 3.4$ &  $1.0 \pm 2.7$ &  $1.8 \pm 2.7$ &  $-0.8 \pm 4.8$ \\
     & \dmsm{} &  $3.9 \pm 2.6$ &  $0.3 \pm 3.5$ &  $2.5 \pm 2.7$ &  $5.5 \pm 3.2$ &  $0.4 \pm 5.0$ \\
    \noalign{\smallskip}
    \hline
    \noalign{\smallskip}
    & Gaussian  &  $-0.3 \pm 3.4$ &  $-0.3 \pm 3.8$ &  $1.6\pm4.1^{\dagger}$ &  $0.0 \pm 2.9$ &  $-1.6 \pm 5.3$ \\
    Baseline rms,       & \dzsz{}     &  $-0.7 \pm 3.4$ &  $-0.06 \pm 3.9$ &  $-0.6 \pm 3.6$ &  $1.1 \pm 3.2$ &  $-1.1 \pm 5.3$ \\
    pessimistic $1/f$   & \doso{}     &  $1.1 \pm 3.4$ &  $-0.6 \pm 4.0$ &  $0.5 \pm 3.6$ &  $3.8 \pm 3.2$ &  $-1.2 \pm 5.3$ \\
    & \dmsm{}     &  $2.8 \pm 3.4$ &  $-0.6 \pm 4.0$ &  $1.2 \pm 3.6$ &  $6.0 \pm 3.1$ &  $-0.5 \pm 5.1$ \\
    \noalign{\smallskip}
    \hline
    \noalign{\smallskip}
    \hline
    \end{tabular}\\
    \footnotesize{$^{\dagger}$ These results are calculated on a smaller, more homogeneous mask, shown in Fig.~\ref{fig:pipelineB_foreground_bias}. This is explained in Appendix~\ref{sec:nilc_gaussian_bias}.}\\
\end{table*}
    
\normalsize

Having presented the results on the CMB power spectra, let us now examine the final constraints on $r$ obtained by each pipeline applied to 500 simulations. These results are summarized in Fig.~ \ref{fig:r_sigma_r_compilation} and Table~ \ref{tab:r_fiducial_results}. Figure~ \ref{fig:r_sigma_r_compilation} shows the mean $r$ and $(16,\,84)$\% credible intervals found by each pipeline as a function of the input foreground model (labels on the $x$ axis). Results are shown for five pipeline setups: pipeline A using the $C_\ell$-fiducial model (red), pipeline A using the $C_\ell$-moments model (yellow), pipeline B (blue), pipeline C (green), and pipeline C including the marginalization over the dust amplitude parameter (cyan). For each pipeline, we show two points with error bars. The dot markers and smaller error bars correspond to the results found in the best-case instrument scenario (goal noise level, optimistic $1/f$ component), while the cross markers and larger error bars correspond to the baseline noise level and pessimistic $1/f$ component. The quantitative results are reported in Table~ \ref{tab:r_fiducial_results}.\\

We start by discussing the nominal pipelines A, B, and C without considering any extensions. We find that for the simpler Gaussian and \dzsz{} foregrounds, the nominal pipelines obtain unbiased results, as expected. Pipeline B shows a slight positive bias for Gaussian foregrounds, in combination with inhomogeneous noise only. This bias is absent for homogeneous noise and can be traced back to the pixel covariance matrix used to construct the NILC weights. We discuss this in more detail in Appendix \ref{sec:nilc_gaussian_bias}. For now, we show the results using a smaller, more homogeneously weighted mask. We stress that these results, marked with a $^{\dagger}$, are not comparable to the rest in Table~\ref{tab:r_fiducial_results}, since they are calculated on a different mask. The more complex \doso{} foregrounds lead to a $\sim1 \sigma$ bias in the goal and optimistic noise scenario. The \dmsm{} foregrounds lead to a noticeable increase of the bias of up to $\sim2\sigma$, seen with pipeline C in all noise scenarios and with pipeline A in the goal-optimistic case, and slightly less with pipeline B. The modifications introduced in the \dmsm{} foreground model include a larger spatial variation in the synchrotron spectral index $\beta_s$ with respect to \doso{}, and are a plausible reason for the increased bias on $r$. 

Remarkably, we find that, in their simplest incarnation, all pipelines achieve comparable statistical uncertainty on $r$, ranging from $\sigma(r)\simeq2.1\times10^{-3}$ to $\sigma(r)\simeq3.6\times10^{-3}$ (a $70\%$ increase), depending on the noise model. Changing between the goal and baseline white noise levels results in an increase of $\sigma(r)$ of $\sim 20-30\%$. Changing between the optimistic and pessimistic $1/f$ noise has a similar effect on the results from pipelines A and B, although $\sigma(r)$ does not increase by more than $10\%$ when changing to pessimistic $1/f$ noise for pipeline C. These results are in reasonable agreement with the forecasts presented in \cite{2019JCAP...02..056A}.
\\

Let us now discuss the pipeline extensions A + moments and C + dust marginalization. Notably, in all noise and foreground scenarios, the two extensions are able to reduce the bias on $r$ to below $1\sigma$. For the Gaussian and \dzsz{} foregrounds, we consistently observe a small negative bias (at the $\sim 0.1\sigma$ level for A + moments and $< 0.5\sigma$ for C + dust marginalization). This bias may be caused by the introduction of extra parameters that are prior dominated, like the dust template’s amplitude in the absence of residual dust contamination, or the moment parameters in the absence of varying spectral indices of foregrounds. If those extra parameters are weakly degenerate with the tensor-to-scalar ratio, the marginal $r$ posterior will shift according to the choice of the prior on the extra parameters. The observed shifts in the tensor-to-scalar ratio and their possible relation with these volume effects will be investigated in a future work. For the more complex \doso{}{} and \dmsm{}, both pipeline extensions effectively remove the bias observed in the nominal pipelines, achieving a $\sim 0.5\sigma$ bias and lower. 

The statistical uncertainty $\sigma(r)$ increases for both pipeline extensions, although by largely different factors. While C + dust marginalization yields $\sigma(r)$ between $3.0\times10^{-3}$ and $5.9\times10^{-3}$, the loss in precision for A + moments is significantly smaller, with $\sigma(r)$ varying between $2.7\times10^{-3}$ and $4.0\times10^{-3}$ depending on the noise scenario, an average increase of $\sim25\%$ compared to pipeline A. In any case, within the assumptions made regarding the SO noise properties, it should be possible to detect a primordial B-mode signal with $r=0.01$ at the 2--3$\sigma$ level with no delensing. The impact of other effects, such as time domain filtering or anisotropic noise may affect these forecasts, and will be studied in more detail in the future.
\\

\begin{figure}
    \centering
    \includegraphics[width=\columnwidth]{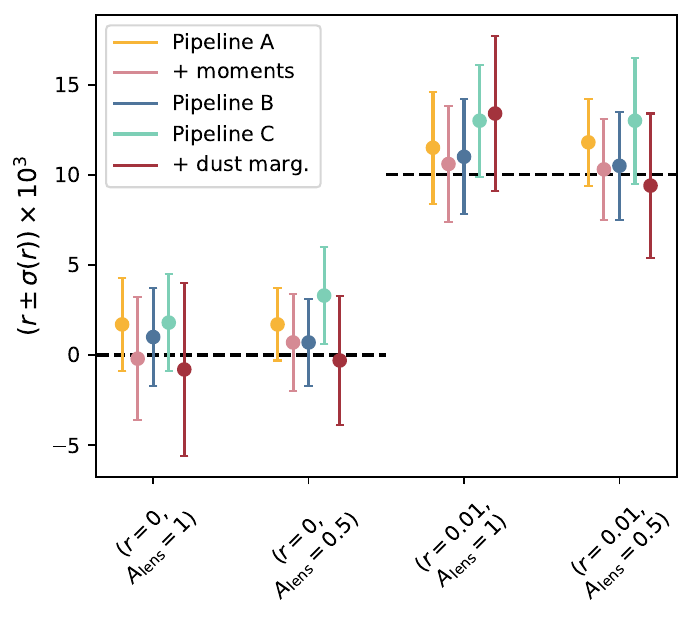}
    \caption{Mean $r$ and $(16,\,84)$\% credible interval from 500 simulations, using the three nominal pipelines plus extensions. We assume input models including primordial B-modes and 50\% delensing efficiency, the SO baseline noise level with optimistic $1/f$ component, and the \doso{} foreground template.}
    \label{fig:r_sigma_r_extra}
\end{figure}

We repeated this analysis for input CMB maps generated assuming either $r=0$ or $0.01$, and either $A_{\rm lens}=0.5$ or $1$. For simplicity, in these cases we considered only the baseline white noise level with optimistic $1/f$ noise and the moderately complex \doso{} foreground model. We show results in Fig.~\ref{fig:r_sigma_r_extra} and Table~\ref{tab:r_extra_results}. A 50\% delensing efficiency results in a reduction in the final $\sigma(r)$ by 25-30\%  for pipelines A and B, $\sim$10-20\% for A + moments, and 0-33\% for C + dust marginalization. The presence of primordial B-modes with a detectable amplitude increases the contribution from cosmic variance to the error budget, with $\sigma(r)$ growing by up to $40\%$ if $r=0.01$, in agreement with theoretical expectations. Using C + dust marginalization and considering no delensing, we even find $\sigma(r)$ decreasing, hinting at the possible breaking of the degeneracy between $r$ and $A_{\rm dust}$. We conclude that all pipelines are able to detect the $r=0.01$ signal at the level of $\sim3\sigma$. As before, we observe a 0.5-1.2$\sigma$ bias on the recovered $r$ that is eliminated by both the moment expansion method and the dust marginalization method. \\

\begin{table*}
\caption{Mean $r$ with $(16,\,84)$\% credible interval from 500 simulations, using the three nominal pipelines with extensions. Our input models contain primordial B-modes of an amplitude $r=0.01$ and 50\% delensing efficiency. We assume the SO baseline noise level with optimistic $1/f$ component and \doso{} foregrounds, see Fig.~\ref{fig:r_sigma_r_extra}.} \label{tab:r_extra_results}
\centering
    \begin{tabular}{cccccc}

    \hline
    \hline
    \noalign{\smallskip}
    \multicolumn{6}{c}{$10^3\times(r\pm\sigma(r))$}\\
    \hline
    \noalign{\smallskip}
    Input CMB & Pipeline A & $+$ moments & Pipeline B & Pipeline C & $+$ dust marg.\\
    \noalign{\smallskip}
    $(r=0,$
    $A_{\rm{lens}}=1)$ &  $1.7 \pm 2.6$ &  $-0.2 \pm 3.4$ &  $1.0 \pm 2.7$ &  $1.8 \pm 2.7$ &  $-0.8 \pm 4.8$ \\
     $(r=0,$
    $A_{\rm{lens}}=0.5)$ &  $1.7 \pm 2.0$ &  $0.7 \pm 2.7$ &  $0.7 \pm 2.4$ &  $3.3 \pm 2.7$ &  $-0.3 \pm 3.6$ \\
     $(r=0.01,$
    $A_{\rm{lens}}=1)$ &  $11.5 \pm 3.1$ &  $10.6 \pm 3.2$ &  $11.0 \pm 3.2$ &  $13.0 \pm 3.1$ &  $13.4 \pm 4.3$ \\
     $(r=0.01,$
    $A_{\rm{lens}}=0.5)$ &  $11.8 \pm 2.4$ &  $10.3 \pm 2.8$ &  $10.5 \pm 3.0$ &  $13.0 \pm 3.5$ &  $9.4 \pm 4.0$ \\
    \hline
    \noalign{\smallskip}
    \hline
    \end{tabular}
\end{table*}

\begin{figure}
    \centering
    \includegraphics[width=\columnwidth]{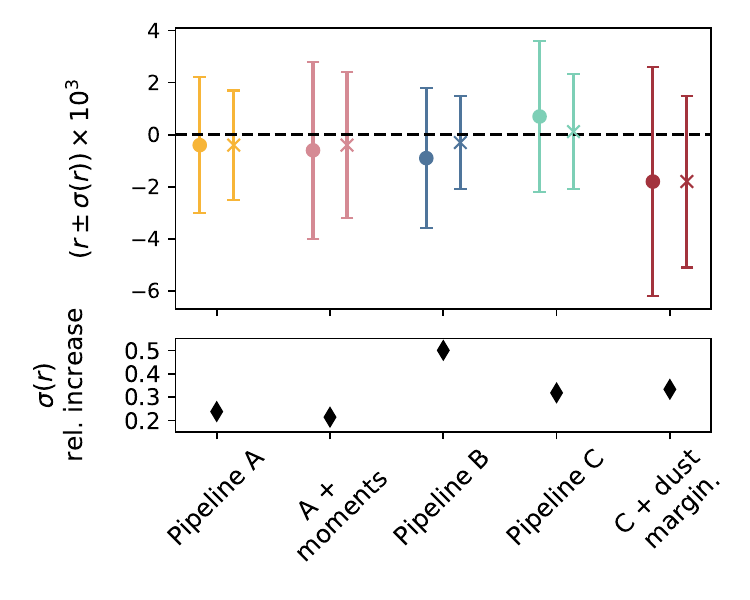}
    \caption{Mean $r$ with $(16,\,84)$\% credible intervals from 500 simulations, applying the three nominal component separation pipelines plus extensions. We assume the \dzsz{} foreground scenario with baseline white noise level and optimistic $1/f$ component. Cross markers with smaller error bars correspond to homogeneous noise across the SAT field of view and dot markers with larger error bars correspond to inhomogeneous noise. The relative increase in $\sigma(r)$ between both is shown in the bottom panel.}
    \label{fig:r_hom_vs_inhom}
\end{figure}

Finally, we explored how cosmological constraints and the pipelines' performances are affected by noise inhomogeneity resulting from weighting the noise pixels according to the SO-SAT hits map. The geographical location of SO and the size of the SAT field of view constrain possible scanning strategies. In particular, SO must target a patch that has a relatively large sky fraction $f_{\rm sky}\sim0.15$ and is surrounded by a $\sim10$ degree wide boundary with significantly higher noise (see hits map in Fig.~\ref{fig:mask}). The lower panel of Fig.~\ref{fig:r_hom_vs_inhom} shows the ratio between the values of $\sigma(r)$ found using inhomogeneous noise realizations and those with homogeneous noise in the baseline-optimistic noise model with \dzsz{} foregrounds, averaged over 500 simulations. We see that for all pipeline scenarios, $\sigma(r)$ increases by $\sim30\%$ due to the noise inhomogeneity.
    
\subsection{Channel weights}\label{ssec:channel_weights}

\begin{figure}
    \centering
    \includegraphics[width=\columnwidth]{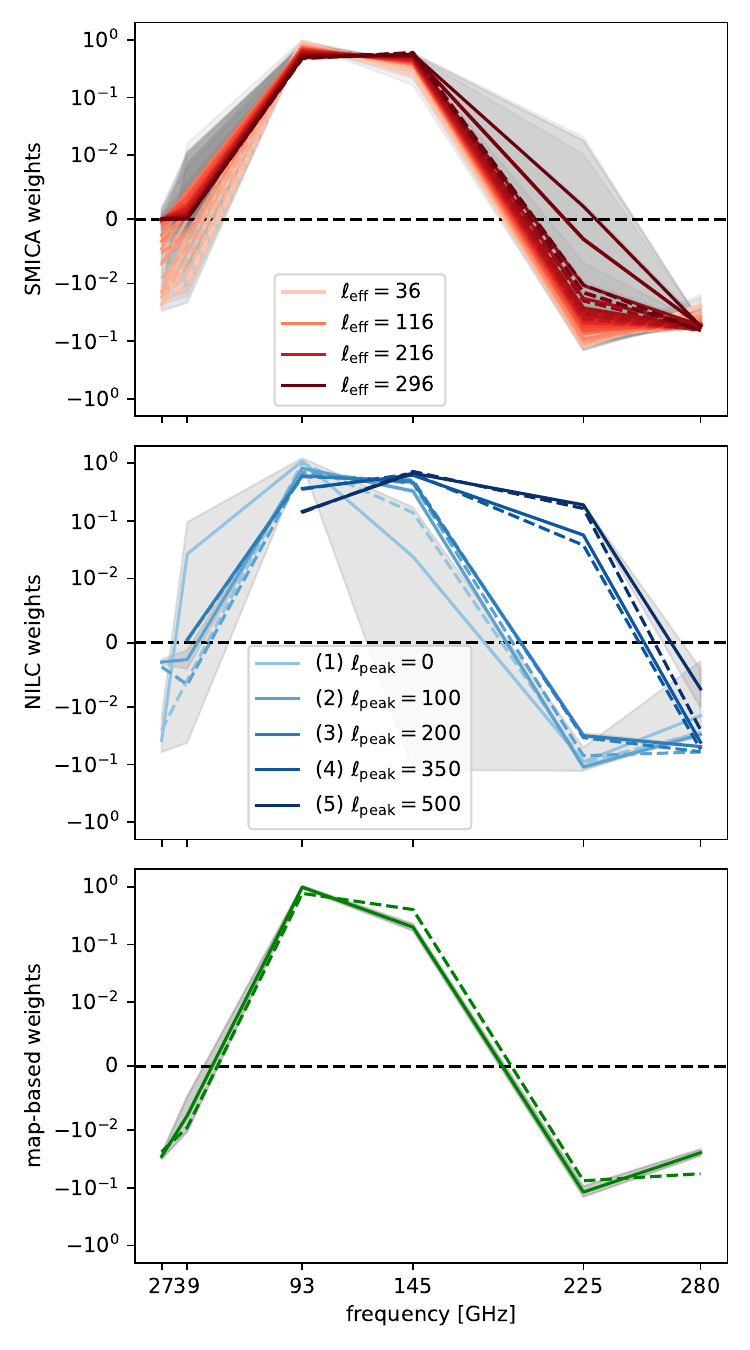}
    \caption{Channel-specific weights associated with the component-separated CMB for the three nominal pipelines. We show the SMICA weights for 27 different $\ell$-bins calculated from raw, noisy $C_\ell$s (pipeline A, upper panel), pixel-averaged NILC weights for five needlet windows (pipeline B, middle panel), and pixel-averaged weights from parametric map-based component separation (pipeline C, lower panel). Weights are averaged over $100$ simulations, shown are goal white $+$ optimistic $1/f$ noise (dashed lines) as well as baseline white $+$ pessimistic $1/f$ noise (solid lines). The semitransparent gray areas represent the channel weights' 1-$\sigma$ standard deviation across $100$ simulations, covering baseline $+$ pessimistic noise.}\label{fig:channel_weights}
\end{figure}
    
Our three baseline pipelines differ fundamentally in how they separate the sky components. One common feature among all pipelines is the use of six frequency channels to distinguish components by means of their different SEDs. In Fig.~\ref{fig:channel_weights} we visualize the channel weights as a function of the band center frequency, showing the pipelines in three vertically stacked panels. In the upper panel, we show the effective weights for the CMB applied to the noise-debiased raw power spectra used by pipeline A, distinguishing between weights for each harmonic bin:
\begin{equation}
    \mathbf{w_{\ell}}^T = \frac{\mathbf{a}^T \hat{\sf C}_{\ell}^{-1} }{\mathbf{a}^T \hat{\sf C}_{\ell}^{-1} \mathbf{a} }.
\end{equation}
Here, $\hat{\sf C}_{\ell}$ is the $6\times 6$ matrix of raw cross-frequency power spectra from noisy sky maps, $\mathbf{a}$ is a vector of length six, filled with ones. This is equivalent to the weights employed by the SMICA component separation method \citep{2008arXiv0803.1814C} and by ILC as explained in Sect.~\ref{sec:nilc}. The middle panel shows the pixel-averaged NILC weights for the five needlet windows (Fig.~\ref{fig:nilc_windows}) used in pipeline B. In the lower panel, we show the CMB weights calculated with the map-based component separation, pipeline C, averaged over the observed pixels to yield an array of six numbers. We averaged all channel weights over 100 simulations containing CMB ($r=0$, $A_{\rm lens}=1$), \doso{} foregrounds, and one of two noise models: goal white noise with optimistic $1/f$ noise is shown as dashed lines, whereas baseline white noise with pessimistic $1/f$ noise is shown as solid lines. Moreover, the gray shaded areas quantify the 1-$\sigma$ uncertainty region of these weights in the baseline-pessimistic case estimated from $100$ simulations. We see from Fig.~\ref{fig:channel_weights} that the average channel weights agree well between pipelines A, B, and C. Mid-frequency channels at $93$ and $145$ GHz are assigned positive CMB weights throughout all pipelines, while high- and low-frequency channels tend to be suppressed owing to a larger dust and synchrotron contamination, respectively. More specifically, the $280$ GHz channel is given negative weight in all pipelines, while average weights at $27$, $39$, and $225$ GHz are negative with pipeline C and either positive or negative in pipelines A and B, depending on the angular scale. The CMB channel weight tends to consistently increase for pipelines A and B as a function of multipole, a fact well exemplified by NILC at $225$ GHz, matching the expectation that the CMB lensing signal becomes more important at high $\ell$. Overall, Fig.~\ref{fig:channel_weights} illustrates that foregrounds at low and high frequencies are consistently subtracted by the three component separation pipelines, with the expected scale dependence in pipelines A and B. Moreover, at every frequency, the channel weights are non-negligible and of similar size across the pipelines, meaning that all channels give a relevant contribution to component separation for all pipelines.

\subsection{More complex foregrounds: \dtsf{}}\label{sec:d10s5}

During the completion of this paper, the new {\sc PySM3} Galactic foreground models\footnote{See \href{https://pysm3.readthedocs.io/en/latest}{pysm3.readthedocs.io/en/latest}.} were made publicly available. In particular, in these new models, templates for polarized thermal dust and synchrotron radiation were updated including the following changes:

\begin{enumerate}
    \item Large-scale thermal dust emission is based on the GNILC maps \citep{2020A&A...641A...4P}, which present a lower contamination from CIB emission with respect to the {\tt d1} model, based on {\tt Commander} templates.
    \item For both thermal dust and synchrotron radiation, small scale structures are added by modifying the logarithm of the polarization fraction tensor\footnote{See \href{https://pysm3.readthedocs.io/en/latest/preprocess-templates/small_scale_dust_pysm3.html}{pysm3.readthedocs.io/en/latest/preprocess-templates}.}.
    \item Thermal dust spectral parameters are based on GNILC products, with larger variation of $\beta_d$ and $T_d$ parameter at low resolution compared to the {\tt d1} model. Small-scale structure is also added as Gaussian realizations of power-law power spectra. 
    \item The new template for $\beta_s$ includes information from the analysis of S-PASS data \citep{2018A&A...618A.166K}, in a similar way as the one of the {\tt sm} model adopted in this work. In addition, small-scale structures are present at sub-degree angular scales.
\end{enumerate}

These modifications are encoded in the models called {\tt d10} and {\tt s5} in the updated version of {\sc PySM}. Although these models are still to be considered preliminary, both in terms of their implementation details in {\sc PySM}\footnote{The {\sc PySM} library is currently under development, with beta versions including minor modifications of the foreground templates being realized regularly. In this part of our analysis we make use of {\sc PySM v3.4.0b3}. } and in general, being based on datasets that may not fully involve the unknown level of foreground complexity, we decided to dedicate an extra section to their analysis. For computational speed, we ran the five pipeline set-ups on a reduced set of 100 simulations containing the new \dtsf{} foregrounds template, CMB with a standard cosmology ($r=0,\, A_{\rm lens}=1$) and inhomogeneous noise in the goal-optimistic scenario. The resulting marginalized posterior mean and $(16,\,84)$\% credible intervals on $r$, averaged over 100 simulations, are:

\begin{align}
    & r\times10^3 = 19.2 \pm 1.9 & (\text{pipeline A})\notag \\
    & r\times10^3 = 2.7 \pm 2.8 & (\text{A + moments})\notag \\
    & r\times10^3 = 15.8 \pm 1.9 & (\text{pipeline B})\\
    & r\times10^3 = 22.0 \pm 2.6 & (\text{pipeline C})\notag \\
    & r\times10^3 = -1.5 \pm 5.1 & (\text{C + dust marg.})\notag
\end{align}

We note that the respective bias obtained with pipelines A, B, and C are at 10, 8, and 8$\sigma$, at least quadrupling the bias of the \dmsm{} foreground model. Crucially, this bias is reduced to less than $1\sigma$ with the A + moments pipeline, with $45\%$ increase in $\sigma(r)$ compared to pipeline A, and $0.3\sigma$ with the C + dust-marginalization pipeline, with a $95\%$ increase in $\sigma(r)$ compared to pipeline C. This makes A + moments the unbiased method with the lowest statistical error.

The $C_\ell$-fiducial model achieves minimum $\chi^2$ values of $601\pm 41$. Although this is an increase of $\Delta\chi^2\sim30$ with respect to the less complex foreground simulations (see Appendix~\ref{app:validation}), the associated probability to exceed (PTE) is $0.10$ (assuming our null distribution is a $\chi^2$ with $N_{\rm data}-N_{\rm parameters}=558$ degrees of freedom), and therefore it would not be possible to identify the presence of a foreground bias by virtue of the model providing a bad fit to the data. The minimum $\chi^2$ values we find also confirm that the covariance matrix calculated from Gaussian simulations is still appropriate for the non-Gaussian \dtsf{} template. On the other hand, A + moments achieves minimum $\chi^2$ values of $537 \pm 33$, which is about $4\%$ lower than for less complex foreground simulations, indicating an improved fitting accuracy. 

As shown in Fig.~\ref{fig:model_odds_d10s5}, the relative model odds between $C_\ell$-fiducial and $C_\ell$-moments (see Appendix~\ref{app:validation} for more details) vary between $10^{-26}$ and $10^{-2}$, clearly favoring $C_\ell$-moments. Out of 100 \dtsf{} simulations, 99 yield model odds below 1\% and 78 below $10^{-5}$. As opposed to the less complex foreground simulations (\dzsz, \doso{}, and \dmsm{}), \dtsf{} gives strong preference to using the moment expansion in the power spectrum model. We note that the AIC-based model odds are computed from the differences of $\chi^2$ values that stem from the same simulation seed and are therefore insensitive to bias from noise and cosmic variance. This explains why AIC odds are the more powerful model comparison tool when compared with the $\chi^2$ analysis presented above.

These results consider only the most optimistic noise scenario. Other cases would likely lead to larger uncertainty and, as a consequence, lower relative biases. In this regard, it is highly encouraging to see two pipeline extensions being able to robustly separate the cosmological signal from Galactic synchrotron and dust emission with this high-level complexity. This highlights the importance of accounting for and marginalizing over residual foreground contamination due to frequency decorrelation for the level of sensitivity that SO and other next-generation observatories will achieve.

The contrast between the results obtained on the \dmsm{} and \dtsf{} simulations gives us an opportunity to reflect on the strategy one should follow when determining the fiducial component separation method to use in primordial B-mode searches. Although the \dmsm{} model leads to a $\sim2\sigma$ bias on $r$ under the simplest component separation algorithms, simple model-selection metrics are not able to provide significant evidence that a more sophisticated modeling of foregrounds is needed. The situation changes with \dtsf{}. A conservative approach is therefore to select the level of complexity needed for component separation by ensuring that unbiased constraints are obtained for all existing foreground models consistent with currently available data. The analysis methods passing this test can then form the basis for the fiducial B-mode constraints. Alternative results can then be obtained with less conservative component separation techniques, but their goodness of fit (or any similar model selection metric) should be compared with that of the fiducial methods. These results should also be accompanied by a comprehensive set of robustness tests able to identify signatures of foreground contamination in the data. This will form the basis of a future work. In a follow-up paper, we will also explore the new set of complex {\sc PySM3} foreground templates in more detail.

\section{Conclusions}\label{sec:conc}

\begin{figure}
    \centering
     \includegraphics[width=\columnwidth]{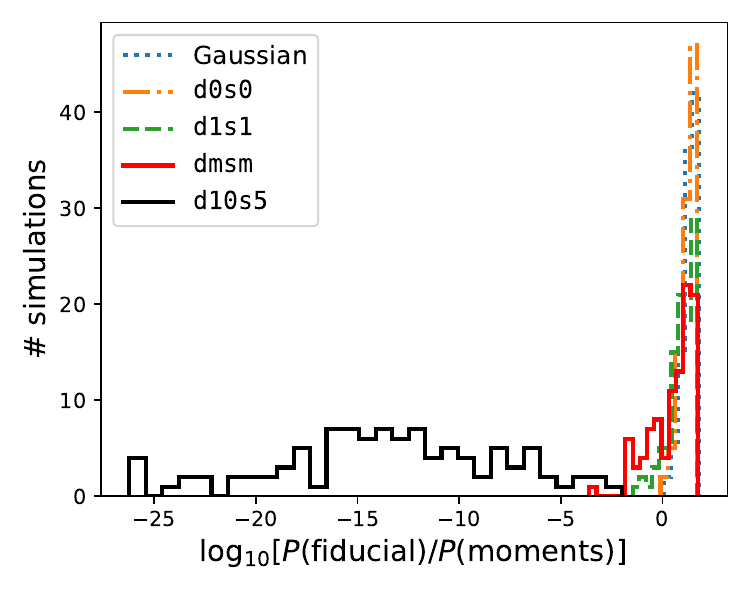}
    \caption{Empirical distribution of the AIC-based relative model odds between the $C_\ell$-fiducial and the $C_\ell$-moments model from 100 simulations. We compare five different Galactic foreground templates, including the {\sc PySM} foreground model \dtsf. Negative values indicate preference for the moments model. We find strong preference for the $C_\ell$-moments model in the \dtsf{} foreground scenario, and only then.} \label{fig:model_odds_d10s5}
\end{figure}

In this paper, we present three different component separation pipelines designed to place constraints on the amplitude of cosmological B-modes on polarized maps of the SO Small Aperture Telescopes. The pipelines are based on multi-frequency $C_\ell$ parametric cleaning (Pipeline A), blind Needlet ILC cleaning (Pipeline B), and map-based parametric cleaning (Pipeline C). We also introduce extensions of pipelines A and C that marginalize over additional residual foreground contamination, using a moment expansion or a dust power spectrum template, respectively. We tested and compared their performance on a set of simulated maps containing lensing B-modes with different scenarios of instrumental noise and Galactic foreground complexity. The presence of additional instrumental complexity, such as time-domain filtering, or anisotropic noise, are likely to affect our results. The impact of these effects will be more thoroughly studied in future work.

We find the inferred uncertainty on the tensor-to-scalar ratio $\sigma(r)$ to be compatible between the three pipelines. While the simpler foreground scenarios (Gaussian, \dzsz{}) do not bias $r$, spectral index variations can cause an increased bias of 1--2$\sigma$ if left untreated, as seen with more complex foreground scenarios (\doso{}, \dmsm{}). Modeling and marginalizing over the spectral residuals is vital to obtain unbiased B-mode estimates. The extensions to pipelines A and C are able to yield unbiased estimates on all foreground scenarios, albeit with a respective increase in $\sigma(r)$ by $\sim 20\%$ (A + moments) and $>30\%$ (C + dust marginalization). These results are in good agreement with the forecasts presented in \cite{2019JCAP...02..056A}.

After testing on simulations with an $r=0.01$ cosmology, we conclude that under realistic conditions and if the forecasted map noise levels and characteristics are achieved, SO should be able to detect a $r=0.01$ signal at $\sim2$--$3\sigma$ after five years of observation. Inhomogeneous noise from the SAT map-making scanning strategy brings about $30\%$ increase in $\sigma(r)$ as compared to homogeneous noise. Analyzing the per-channel weights for our pipelines, we find all frequency channels to be relevant for the CMB signal extraction and all pipelines to be in good agreement. These forecasts cover the nominal SO survey, and can be considered pessimistic in the light of prospective additional SATs that will further improve the sensitivity on large angular scales.

We also carried out a preliminary analysis of new, more complex, foreground models recently implemented in {\sc PySM3}, in particular the \dtsf{} foreground template. The much higher level of spatial SED variation allowed by this model leads to a drastic increase in the bias on $r$ by up to $10\sigma$, when analyzed with the nominal pipelines A, B, and C. Fortunately, this bias can be reduced to below $1\sigma$ when using A + moments and C + dust marginalization. These extensions lead to a 45\% and 95\% degradation of the error bars, respectively. Our results highlight the importance of marginalizing over residuals caused by frequency decorrelation for SO-like sensitivities. Although our analysis of \dtsf{} is less exhaustive than that of the other foreground models presented here, it is encouraging to confirm that we have the tools at hand to obtain robust, unbiased constraints on the tensor-to-scalar ratio in the presence of such complex Galactic foregrounds. In addition to the algorithmic improvements presented in this paper, the inclusion of external data sets such as FYST/CCAT-Prime \citep{CCAT-Prime_2023} may prove helpful at mitigating foregrounds.

In preparation for the data collected by SO in the near future, we will continue our investigations into Galactic foreground models with other levels of complexity as the field progresses. Nevertheless, the current work shows that the analysis pipelines in place for SO are able to obtain robust constraints on the amplitude of primordial $B$ modes in the presence of Galactic foregrounds covering the full range of complexity envisaged by current, state-of-the-art models.

\begin{acknowledgements}

The authors would like to thank Ken Ganga, Arthur Kosowsky, and the anonymous referee for useful feedback. The group at SISSA acknowledges support from the COSMOS Network of the Italian Space Agency and the InDark Initiative of the National Institute for Nuclear Physics (INFN). 
KW is funded by a SISSA PhD fellowship. SA is funded by a Kavli/IPMU doctoral studentship. 
CHC acknowledges NSF award 1815887 and FONDECYT Postdoc fellowship 3220255. 
DA is supported by the Science and Technology Facilities Council through an Ernest Rutherford Fellowship, grant reference ST/P004474.
This work is part of a project that has received funding from the
European Research Council (ERC) under the European Union’s Horizon 2020 research and innovation program (PI: Josquin Errard, Grant agreement No. 101044073). 
ABL is a BCCP fellow at UC Berkeley and Lawrence Berkeley National Laboratory.
MLB acknowledges funding from UKRI and STFC (Grant awards ST/X006344/1 and ST/X006336/1).
EC acknowledges support from the European Research Council (ERC) under the European Union’s Horizon 2020 research and innovation programme (Grant agreement No. 849169).
JC was furthermore supported by the ERC Consolidator Grant {\it CMBSPEC} (No.~725456) and the Royal Society as a Royal Society University Research Fellow at the University of Manchester, UK (No.~URF/R/191023).
GF acknowledges the support of the European Research Council under the Marie Sk\l{}odowska Curie actions through the Individual Global Fellowship No.~892401 PiCOGAMBAS.

We acknowledge the use of  \texttt{CAMB} \citep{Lewis:1999bs}, \texttt{healpy} \citep{Zonca2019}, \texttt{numpy} \citep{Harris:2020xlr}, \texttt{matplotlib} \citep{Hunter:2007ouj}, \texttt{emcee} \citep{2013PASP..125..306F}, and \texttt{fgbuster} \citep{Errard2019,Puglisi2022} software packages.

\end{acknowledgements}

\bibliographystyle{aa}
\bibliography{bibliography}

\begin{thebibliography}{84}
\expandafter\ifx\csname natexlab\endcsname\relax\def\natexlab#1{#1}\fi

\bibitem[{{Abazajian} {et~al.}(2016){Abazajian}, {Adshead}, {Ahmed}, {Allen},
  {Alonso}, {Arnold}, {Baccigalupi}, {Bartlett}, {Battaglia}, {Benson},
  {Bischoff}, {Borrill}, {Buza}, {Calabrese}, {Caldwell}, {Carlstrom}, {Chang},
  {Crawford}, {Cyr-Racine}, {De Bernardis}, {de Haan}, {di Serego Alighieri},
  {Dunkley}, {Dvorkin}, {Errard}, {Fabbian}, {Feeney}, {Ferraro}, {Filippini},
  {Flauger}, {Fuller}, {Gluscevic}, {Green}, {Grin}, {Grohs}, {Henning},
  {Hill}, {Hlozek}, {Holder}, {Holzapfel}, {Hu}, {Huffenberger}, {Keskitalo},
  {Knox}, {Kosowsky}, {Kovac}, {Kovetz}, {Kuo}, {Kusaka}, {Le Jeune}, {Lee},
  {Lilley}, {Loverde}, {Madhavacheril}, {Mantz}, {Marsh}, {McMahon},
  {Meerburg}, {Meyers}, {Miller}, {Munoz}, {Nguyen}, {Niemack}, {Peloso},
  {Peloton}, {Pogosian}, {Pryke}, {Raveri}, {Reichardt}, {Rocha}, {Rotti},
  {Schaan}, {Schmittfull}, {Scott}, {Sehgal}, {Shandera}, {Sherwin}, {Smith},
  {Sorbo}, {Starkman}, {Story}, {van Engelen}, {Vieira}, {Watson}, {Whitehorn},
  \& {Kimmy Wu}}]{2016arXiv161002743A}
{Abazajian}, K.~N., {Adshead}, P., {Ahmed}, Z., {et~al.} 2016, arXiv e-prints,
  arXiv:1610.02743

\bibitem[{{Abbott} \& {Wise}(1984)}]{1984NuPhB.244..541A}
{Abbott}, L.~F. \& {Wise}, M.~B. 1984, Nuclear Physics B, 244, 541

\bibitem[{{Abitbol} {et~al.}(2021){Abitbol}, {Alonso}, {Simon}, {Lashner},
  {Crowley}, {Ali}, {Azzoni}, {Baccigalupi}, {Barron}, {Brown}, {Calabrese},
  {Carron}, {Chinone}, {Chluba}, {Coppi}, {Crowley}, {Devlin}, {Dunkley},
  {Errard}, {Fanfani}, {Galitzki}, {Gerbino}, {Hill}, {Johnson}, {Jost},
  {Keating}, {Krachmalnicoff}, {Kusaka}, {Lee}, {Louis}, {Madhavacheril},
  {McCarrick}, {McMahon}, {Meerburg}, {Nati}, {Nishino}, {Page}, {Poletti},
  {Puglisi}, {Randall}, {Rotti}, {Spisak}, {Suzuki}, {Teply}, {Verg{\`e}s},
  {Wollack}, {Xu}, \& {Zannoni}}]{2021JCAP...05..032A}
{Abitbol}, M.~H., {Alonso}, D., {Simon}, S.~M., {et~al.} 2021, \jcap, 2021, 032

\bibitem[{Akaike(1974)}]{1100705}
Akaike, H. 1974, IEEE Transactions on Automatic Control, 19, 716

\bibitem[{{Alonso} {et~al.}(2017){Alonso}, {Dunkley}, {Thorne}, \&
  {N{\ae}ss}}]{2017PhRvD..95d3504A}
{Alonso}, D., {Dunkley}, J., {Thorne}, B., \& {N{\ae}ss}, S. 2017, \prd, 95,
  043504

\bibitem[{{Alonso} {et~al.}(2019){Alonso}, {Sanchez}, {Slosar}, \& {LSST Dark
  Energy Science Collaboration}}]{2019MNRAS.484.4127A}
{Alonso}, D., {Sanchez}, J., {Slosar}, A., \& {LSST Dark Energy Science
  Collaboration}. 2019, \mnras, 484, 4127

\bibitem[{{Armitage-Caplan} {et~al.}(2012){Armitage-Caplan}, {Dunkley},
  {Eriksen}, \& {Dickinson}}]{2012MNRAS.424.1914A}
{Armitage-Caplan}, C., {Dunkley}, J., {Eriksen}, H.~K., \& {Dickinson}, C.
  2012, \mnras, 424, 1914

\bibitem[{{Azzoni} {et~al.}(2021){Azzoni}, {Abitbol}, {Alonso}, {Gough},
  {Katayama}, \& {Matsumura}}]{2021JCAP...05..047A}
{Azzoni}, S., {Abitbol}, M.~H., {Alonso}, D., {et~al.} 2021, \jcap, 2021, 047

\bibitem[{{Basak} \& {Delabrouille}(2012)}]{2012MNRAS.419.1163B}
{Basak}, S. \& {Delabrouille}, J. 2012, \mnras, 419, 1163

\bibitem[{{Basak} \& {Delabrouille}(2013)}]{2013MNRAS.435...18B}
{Basak}, S. \& {Delabrouille}, J. 2013, \mnras, 435, 18

\bibitem[{{Bennett} {et~al.}(2003){Bennett}, {Hill}, {Hinshaw}, {Nolta},
  {Odegard}, {Page}, {Spergel}, {Weiland}, {Wright}, {Halpern}, {Jarosik},
  {Kogut}, {Limon}, {Meyer}, {Tucker}, \& {Wollack}}]{2003ApJS..148...97B}
{Bennett}, C.~L., {Hill}, R.~S., {Hinshaw}, G., {et~al.} 2003, \apjs, 148, 97

\bibitem[{{Betoule} {et~al.}(2009){Betoule}, {Pierpaoli}, {Delabrouille}, {Le
  Jeune}, \& {Cardoso}}]{2009A&A...503..691B}
{Betoule}, M., {Pierpaoli}, E., {Delabrouille}, J., {Le Jeune}, M., \&
  {Cardoso}, J.~F. 2009, \aap, 503, 691

\bibitem[{{BICEP2 Collaboration} \& {Keck Array
  Collaboration}(2016)}]{2016PhRvL.116c1302B}
{BICEP2 Collaboration} \& {Keck Array Collaboration}. 2016, \prl, 116, 031302

\bibitem[{{BICEP2 Collaboration} \& {Keck Array
  Collaboration}(2018)}]{2018PhRvL.121v1301B}
{BICEP2 Collaboration} \& {Keck Array Collaboration}. 2018, \prl, 121, 221301

\bibitem[{{BICEP/Keck Collaboration}(2021)}]{2021PhRvL.127o1301A}
{BICEP/Keck Collaboration}. 2021, \prl, 127, 151301

\bibitem[{{Bonaldi} \& {Ricciardi}(2011)}]{2011MNRAS.414..615B}
{Bonaldi}, A. \& {Ricciardi}, S. 2011, \mnras, 414, 615

\bibitem[{{Cardoso} {et~al.}(2008){Cardoso}, {Martin}, {Delabrouille},
  {Betoule}, \& {Patanchon}}]{2008arXiv0803.1814C}
{Cardoso}, J.-F., {Martin}, M., {Delabrouille}, J., {Betoule}, M., \&
  {Patanchon}, G. 2008, arXiv e-prints, arXiv:0803.1814

\bibitem[{{CCAT-Prime Collaboration} {et~al.}(2022){CCAT-Prime Collaboration},
  Aravena, Austermann, Basu, Battaglia, Beringue, Bertoldi, Bigiel, Bond,
  Breysse, Broughton, Bustos, Chapman, Charmetant, Choi, Chung, Clark, Cothard,
  Crites, Dev, Douglas, Duell, Dünner, Ebina, Erler, Fich, Fissel, Foreman,
  Freundt, Gallardo, Gao, García, Giovanelli, Golec, Groppi, Haynes, Henke,
  Hensley, Herter, Higgins, Hložek, Huber, Huber, Hubmayr, Jackson, Johnstone,
  Karoumpis, Keating, Komatsu, Li, Magnelli, Matthews, Mauskopf, McMahon,
  Meerburg, Meyers, Muralidhara, Murray, Niemack, Nikola, Okada, Puddu,
  Riechers, Rosolowsky, Rossi, Rotermund, Roy, Sadavoy, Schaaf, Schilke, Scott,
  Simon, Sinclair, Sivakoff, Stacey, Stutz, Stutzki, Tahani, Thanjavur,
  Timmermann, Ullom, van Engelen, Vavagiakis, Vissers, Wheeler, White, Zhu, \&
  Zou}]{CCAT-Prime_2023}
{CCAT-Prime Collaboration}, Aravena, M., Austermann, J.~E., {et~al.} 2022, The
  Astrophysical Journal Supplement Series, 264, 7

\bibitem[{{CMB-S4 Collaboration}(2022)}]{2022ApJ...926...54A}
{CMB-S4 Collaboration}. 2022, \apj, 926, 54

\bibitem[{Delabrouille \& Cardoso(2007)}]{Delabrouille:2007bq}
Delabrouille, J. \& Cardoso, J.~F. 2007, in {International Summer School on
  Data Analysis in Cosmology}

\bibitem[{{Delabrouille} {et~al.}(2009){Delabrouille}, {Cardoso}, {Le Jeune},
  {Betoule}, {Fay}, \& {Guilloux}}]{2009A&A...493..835D}
{Delabrouille}, J., {Cardoso}, J.~F., {Le Jeune}, M., {et~al.} 2009, \aap, 493,
  835

\bibitem[{Dunkley {et~al.}(2013)}]{Dunkley:2013vu}
Dunkley, J. {et~al.} 2013, JCAP, 07, 025

\bibitem[{{Errard} {et~al.}(2016){Errard}, {Feeney}, {Peiris}, \&
  {Jaffe}}]{2016JCAP...03..052E}
{Errard}, J., {Feeney}, S.~M., {Peiris}, H.~V., \& {Jaffe}, A.~H. 2016, \jcap,
  2016, 052

\bibitem[{{Errard} \& {Stompor}(2012)}]{2012PhRvD..85h3006E}
{Errard}, J. \& {Stompor}, R. 2012, \prd, 85, 083006

\bibitem[{{Errard} \& {Stompor}(2019)}]{Errard2019}
{Errard}, J. \& {Stompor}, R. 2019, \prd, 99, 043529

\bibitem[{{Foreman-Mackey} {et~al.}(2013){Foreman-Mackey}, {Hogg}, {Lang}, \&
  {Goodman}}]{2013PASP..125..306F}
{Foreman-Mackey}, D., {Hogg}, D.~W., {Lang}, D., \& {Goodman}, J. 2013, \pasp,
  125, 306

\bibitem[{{G{\'o}rski} {et~al.}(2005){G{\'o}rski}, {Hivon}, {Banday},
  {Wandelt}, {Hansen}, {Reinecke}, \& {Bartelmann}}]{2005ApJ...622..759G}
{G{\'o}rski}, K.~M., {Hivon}, E., {Banday}, A.~J., {et~al.} 2005, \apj, 622,
  759

\bibitem[{{Grain} {et~al.}(2009){Grain}, {Tristram}, \&
  {Stompor}}]{2009PhRvD..79l3515G}
{Grain}, J., {Tristram}, M., \& {Stompor}, R. 2009, \prd, 79, 123515

\bibitem[{{Hamimeche} \& {Lewis}(2008)}]{hamimechelewis2008}
{Hamimeche}, S. \& {Lewis}, A. 2008, \prd, 77, 103013

\bibitem[{{Harper} {et~al.}(2022){Harper}, {Dickinson}, {Barr},
  {Cepeda-Arroita}, {Grumitt}, {Heilgendorff}, {Jew}, {Jonas}, {Jones},
  {Leahy}, {Leech}, {Pearson}, {Peel}, {Readhead}, \&
  {Taylor}}]{2022MNRAS.513.5900H}
{Harper}, S.~E., {Dickinson}, C., {Barr}, A., {et~al.} 2022, \mnras, 513, 5900

\bibitem[{Harris {et~al.}(2020)}]{Harris:2020xlr}
Harris, C.~R. {et~al.} 2020, Nature, 585, 357

\bibitem[{{Hazumi} {et~al.}(2019){Hazumi}, {Ade}, {Akiba}, {Alonso}, {Arnold},
  {Aumont}, {Baccigalupi}, {Barron}, {Basak}, {Beckman}, {Borrill},
  {Boulanger}, {Bucher}, {Calabrese}, {Chinone}, {Cho}, {Cukierman}, {Curtis},
  {de Haan}, {Dobbs}, {Dominjon}, {Dotani}, {Duband}, {Ducout}, {Dunkley},
  {Duval}, {Elleflot}, {Eriksen}, {Errard}, {Fischer}, {Fujino}, {Funaki},
  {Fuskeland}, {Ganga}, {Goeckner-Wald}, {Grain}, {Halverson}, {Hamada},
  {Hasebe}, {Hasegawa}, {Hattori}, {Hattori}, {Hayes}, {Hidehira}, {Hill},
  {Hilton}, {Hubmayr}, {Ichiki}, {Iida}, {Imada}, {Inoue}, {Inoue}, {Irwin},
  {Ishino}, {Jeong}, {Kanai}, {Kaneko}, {Kashima}, {Katayama}, {Kawasaki},
  {Kernasovskiy}, {Keskitalo}, {Kibayashi}, {Kida}, {Kimura}, {Kisner},
  {Kohri}, {Komatsu}, {Komatsu}, {Kuo}, {Kurinsky}, {Kusaka}, {Lazarian},
  {Lee}, {Li}, {Linder}, {Maffei}, {Mangilli}, {Maki}, {Matsumura}, {Matsuura},
  {Meilhan}, {Mima}, {Minami}, {Mitsuda}, {Montier}, {Nagai}, {Nagasaki},
  {Nagata}, {Nakajima}, {Nakamura}, {Namikawa}, {Naruse}, {Nishino}, {Nitta},
  {Noguchi}, {Ogawa}, {Oguri}, {Okada}, {Okamoto}, {Okamura}, {Otani},
  {Patanchon}, {Pisano}, {Rebeiz}, {Remazeilles}, {Richards}, {Sakai},
  {Sakurai}, {Sato}, {Sato}, {Sawada}, {Segawa}, {Sekimoto}, {Seljak},
  {Sherwin}, {Shimizu}, {Shinozaki}, {Stompor}, {Sugai}, {Sugita}, {Suzuki},
  {Suzuki}, {Tajima}, {Takada}, {Takaku}, {Takakura}, {Takatori}, {Tanabe},
  {Taylor}, {Thompson}, {Thorne}, {Tomaru}, {Tomida}, {Tomita}, {Tristram},
  {Tucker}, {Turin}, {Tsujimoto}, {Uozumi}, {Utsunomiya}, {Uzawa}, {Vansyngel},
  {Wehus}, {Westbrook}, {Willer}, {Whitehorn}, {Yamada}, {Yamamoto},
  {Yamasaki}, {Yamashita}, \& {Yoshida}}]{2019JLTP..194..443H}
{Hazumi}, M., {Ade}, P.~A.~R., {Akiba}, Y., {et~al.} 2019, Journal of Low
  Temperature Physics, 194, 443

\bibitem[{{Herv{\'\i}as-Caimapo} {et~al.}(2017){Herv{\'\i}as-Caimapo},
  {Bonaldi}, \& {Brown}}]{2017MNRAS.468.4408H}
{Herv{\'\i}as-Caimapo}, C., {Bonaldi}, A., \& {Brown}, M.~L. 2017, \mnras, 468,
  4408

\bibitem[{{Herv{\'\i}as-Caimapo} {et~al.}(2022){Herv{\'\i}as-Caimapo},
  {Bonaldi}, {Brown}, \& {Huffenberger}}]{2022ApJ...924...11H}
{Herv{\'\i}as-Caimapo}, C., {Bonaldi}, A., {Brown}, M.~L., \& {Huffenberger},
  K.~M. 2022, \apj, 924, 11

\bibitem[{{Hui} {et~al.}(2018){Hui}, {Ade}, {Ahmed}, {Aikin}, {Alexander},
  {Barkats}, {Benton}, {Bischoff}, {Bock}, {Bowens-Rubin}, {Brevik}, {Buder},
  {Bullock}, {Buza}, {Connors}, {Cornelison}, {Crill}, {Crumrine}, {Dierickx},
  {Duband}, {Dvorkin}, {Filippini}, {Fliescher}, {Grayson}, {Hall}, {Halpern},
  {Harrison}, {Hildebrandt}, {Hilton}, {Irwin}, {Kang}, {Karkare}, {Karpel},
  {Kaufman}, {Keating}, {Kefeli}, {Kernasovskiy}, {Kovac}, {Kuo}, {Lau},
  {Larsen}, {Leitch}, {Lueker}, {Megerian}, {Moncelsi}, {Namikawa},
  {Netterfield}, {Nguyen}, {O'Brient}, {Ogburn}, {Palladino}, {Pryke},
  {Racine}, {Richter}, {Schwarz}, {Schillaci}, {Sheehy}, {Soliman}, {St.
  Germaine}, {Staniszewski}, {Steinbach}, {Sudiwala}, {Teply}, {Thompson},
  {Tolan}, {Tucker}, {Turner}, {Umilt{\`a}}, {Vieregg}, {Wandui}, {Weber},
  {Wiebe}, {Willmert}, {Wong}, {Wu}, {Yang}, {Yoon}, \& {Zhang}}]{biceparr}
{Hui}, H., {Ade}, P.~A.~R., {Ahmed}, Z., {et~al.} 2018, Proc. SPIE Int. Soc.
  Opt. Eng., 10708 [\eprint[arXiv]{1808.00568}]

\bibitem[{Hunter(2007)}]{Hunter:2007ouj}
Hunter, J.~D. 2007, Comput. Sci. Eng., 9, 90

\bibitem[{{Ijjas} \& {Steinhardt}(2018)}]{2018CQGra..35m5004I}
{Ijjas}, A. \& {Steinhardt}, P.~J. 2018, Classical and Quantum Gravity, 35,
  135004

\bibitem[{{Ijjas} \& {Steinhardt}(2019)}]{2019PhLB..795..666I}
{Ijjas}, A. \& {Steinhardt}, P.~J. 2019, Physics Letters B, 795, 666

\bibitem[{{Kamionkowski} {et~al.}(1997){Kamionkowski}, {Kosowsky}, \&
  {Stebbins}}]{1997PhRvL..78.2058K}
{Kamionkowski}, M., {Kosowsky}, A., \& {Stebbins}, A. 1997, \prl, 78, 2058

\bibitem[{{Kamionkowski} \& {Kovetz}(2016)}]{2016ARA&A..54..227K}
{Kamionkowski}, M. \& {Kovetz}, E.~D. 2016, \araa, 54, 227

\bibitem[{{Katayama} \& {Komatsu}(2011)}]{2011ApJ...737...78K}
{Katayama}, N. \& {Komatsu}, E. 2011, \apj, 737, 78

\bibitem[{{Krachmalnicoff} {et~al.}(2016){Krachmalnicoff}, {Baccigalupi},
  {Aumont}, {Bersanelli}, \& {Mennella}}]{2016A&A...588A..65K}
{Krachmalnicoff}, N., {Baccigalupi}, C., {Aumont}, J., {Bersanelli}, M., \&
  {Mennella}, A. 2016, \aap, 588, A65

\bibitem[{{Krachmalnicoff} {et~al.}(2018){Krachmalnicoff}, {Carretti},
  {Baccigalupi}, {Bernardi}, {Brown}, {Gaensler}, {Haverkorn}, {Kesteven},
  {Perrotta}, {Poppi}, \& {Staveley-Smith}}]{2018A&A...618A.166K}
{Krachmalnicoff}, N., {Carretti}, E., {Baccigalupi}, C., {et~al.} 2018, \aap,
  618, A166

\bibitem[{{Leach} {et~al.}(2008){Leach}, {Cardoso}, {Baccigalupi}, {Barreiro},
  {Betoule}, {Bobin}, {Bonaldi}, {Delabrouille}, {de Zotti}, {Dickinson},
  {Eriksen}, {Gonz{\'a}lez-Nuevo}, {Hansen}, {Herranz}, {Le Jeune},
  {L{\'o}pez-Caniego}, {Mart{\'\i}nez-Gonz{\'a}lez}, {Massardi}, {Melin},
  {Miville-Desch{\^e}nes}, {Patanchon}, {Prunet}, {Ricciardi}, {Salerno},
  {Sanz}, {Starck}, {Stivoli}, {Stolyarov}, {Stompor}, \&
  {Vielva}}]{2008A&A...491..597L}
{Leach}, S.~M., {Cardoso}, J.~F., {Baccigalupi}, C., {et~al.} 2008, \aap, 491,
  597

\bibitem[{{Lewis} \& {Challinor}(2006)}]{2006PhR...429....1L}
{Lewis}, A. \& {Challinor}, A. 2006, \physrep, 429, 1

\bibitem[{Lewis {et~al.}(2000)Lewis, Challinor, \& Lasenby}]{Lewis:1999bs}
Lewis, A., Challinor, A., \& Lasenby, A. 2000, \apj, 538, 473

\bibitem[{{LiteBIRD Collaboration}(2022)}]{PTEP2022}
{LiteBIRD Collaboration}. 2022, Progress of Theoretical and Experimental
  Physics, ptac150

\bibitem[{Maltoni \& Schwetz(2003)}]{Maltoni:2003cu}
Maltoni, M. \& Schwetz, T. 2003, Phys. Rev. D, 68, 033020

\bibitem[{{Martin} {et~al.}(2014{\natexlab{a}}){Martin}, {Ringeval}, {Trotta},
  \& {Vennin}}]{2014JCAP...03..039M}
{Martin}, J., {Ringeval}, C., {Trotta}, R., \& {Vennin}, V. 2014{\natexlab{a}},
  \jcap, 2014, 039

\bibitem[{{Martin} {et~al.}(2014{\natexlab{b}}){Martin}, {Ringeval}, \&
  {Vennin}}]{2014PDU.....5...75M}
{Martin}, J., {Ringeval}, C., \& {Vennin}, V. 2014{\natexlab{b}}, Physics of
  the Dark Universe, 5, 75

\bibitem[{{Namikawa} {et~al.}(2022){Namikawa}, {Baleato Lizancos}, {Robertson},
  {Sherwin}, {Challinor}, {Alonso}, {Azzoni}, {Baccigalupi}, {Calabrese},
  {Carron}, {Chinone}, {Chluba}, {Coppi}, {Errard}, {Fabbian}, {Ferraro},
  {Kalaja}, {Lewis}, {Madhavacheril}, {Meerburg}, {Meyers}, {Nati}, {Orlando},
  {Poletti}, {Puglisi}, {Remazeilles}, {Sehgal}, {Tajima}, {Teply}, {van
  Engelen}, {Wollack}, {Xu}, {Yu}, {Zhu}, \& {Zonca}}]{2022PhRvD.105b3511N}
{Namikawa}, T., {Baleato Lizancos}, A., {Robertson}, N., {et~al.} 2022, \prd,
  105, 023511

\bibitem[{{Nash}(1984)}]{1984SJNA...21..770N}
{Nash}, S.~G. 1984, SIAM Journal on Numerical Analysis, 21, 770

\bibitem[{{Natoli} {et~al.}(2018){Natoli}, {Ashdown}, {Banerji}, {Borrill},
  {Buzzelli}, {de Gasperis}, {Delabrouille}, {Hivon}, {Molinari}, {Patanchon},
  {Polastri}, {Tomasi}, {Bouchet}, {Henrot-Versill{\'e}}, {Hoang}, {Keskitalo},
  {Kiiveri}, {Kisner}, {Lindholm}, {McCarthy}, {Piacentini}, {Perdereau},
  {Polenta}, {Tristram}, {Achucarro}, {Ade}, {Allison}, {Baccigalupi},
  {Ballardini}, {Banday}, {Bartlett}, {Bartolo}, {Basak}, {Baumann},
  {Bersanelli}, {Bonaldi}, {Bonato}, {Boulanger}, {Brinckmann}, {Bucher},
  {Burigana}, {Cai}, {Calvo}, {Carvalho}, {Castellano}, {Challinor}, {Chluba},
  {Clesse}, {Colantoni}, {Coppolecchia}, {Crook}, {D'Alessandro}, {de
  Bernardis}, {De Zotti}, {Di Valentino}, {Diego}, {Errard}, {Feeney},
  {Fernandez-Cobos}, {Finelli}, {Forastieri}, {Galli}, {Genova-Santos},
  {Gerbino}, {Gonz{\'a}lez-Nuevo}, {Grandis}, {Greenslade}, {Gruppuso},
  {Hagstotz}, {Hanany}, {Handley}, {Hernandez-Monteagudo},
  {Herv{\'\i}as-Caimapo}, {Hills}, {Keih{\"a}nen}, {Kitching}, {Kunz},
  {Kurki-Suonio}, {Lamagna}, {Lasenby}, {Lattanzi}, {Lesgourgues}, {Lewis},
  {Liguori}, {L{\'o}pez-Caniego}, {Luzzi}, {Maffei}, {Mandolesi},
  {Martinez-Gonz{\'a}lez}, {Martins}, {Masi}, {Matarrese}, {Melchiorri},
  {Melin}, {Migliaccio}, {Monfardini}, {Negrello}, {Notari}, {Pagano},
  {Paiella}, {Paoletti}, {Piat}, {Pisano}, {Pollo}, {Poulin}, {Quartin},
  {Remazeilles}, {Roman}, {Rossi}, {Rubino-Martin}, {Salvati}, {Signorelli},
  {Tartari}, {Tramonte}, {Trappe}, {Trombetti}, {Tucker}, {Valiviita}, {Van de
  Weijgaert}, {van Tent}, {Vennin}, {Vielva}, {Vittorio}, {Wallis}, {Young}, \&
  {Zannoni}}]{2018JCAP...04..022N}
{Natoli}, P., {Ashdown}, M., {Banerji}, R., {et~al.} 2018, \jcap, 2018, 022

\bibitem[{{Pearson}(1900)}]{pearson1900}
{Pearson}, K. 1900, The London, Edinburgh, and Dublin Philosophical Magazine
  and Journal of Science, 50, 157

\bibitem[{{Planck Collaboration Int. XXX}(2016)}]{2016A&A...586A.133P}
{Planck Collaboration Int. XXX}. 2016, \aap, 586, A133

\bibitem[{{Planck Collaboration IV}(2020)}]{2020A&A...641A...4P}
{Planck Collaboration IV}. 2020, \aap, 641, A4

\bibitem[{{Planck Collaboration V}(2020)}]{planck_v_2020}
{Planck Collaboration V}. 2020, A\&A, 641, A5

\bibitem[{{Planck Collaboration VI}(2020)}]{2020A&A...641A...6P}
{Planck Collaboration VI}. 2020, \aap, 641, A6

\bibitem[{{Planck Collaboration X}(2016)}]{2016A&A...594A..10P}
{Planck Collaboration X}. 2016, \aap, 594, A10

\bibitem[{{Planck Collaboration X}(2020)}]{2020A&A...641A..10P}
{Planck Collaboration X}. 2020, \aap, 641, A10

\bibitem[{{Planck Collaboration XI}(2016)}]{planck_xi_2015}
{Planck Collaboration XI}. 2016, Astron. Astrophys., 594, A11

\bibitem[{{Planck Collaboration XI}(2020)}]{2020A&A...641A..11P}
{Planck Collaboration XI}. 2020, \aap, 641, A11

\bibitem[{{Poletti} \& {Errard}(2023)}]{Poletti_Errard_2022}
{Poletti}, D. \& {Errard}, J. 2023, In prep.

\bibitem[{{Puglisi} {et~al.}(2022){Puglisi}, {Mihaylov}, {Panopoulou},
  {Poletti}, {Errard}, {Puglisi}, \& {Vianello}}]{Puglisi2022}
{Puglisi}, G., {Mihaylov}, G., {Panopoulou}, G.~V., {et~al.} 2022, \mnras, 511,
  2052

\bibitem[{{Remazeilles} {et~al.}(2018{\natexlab{a}}){Remazeilles}, {Banday},
  {Baccigalupi}, {Basak}, {Bonaldi}, {De Zotti}, {Delabrouille}, {Dickinson},
  {Eriksen}, {Errard}, {Fernandez-Cobos}, {Fuskeland}, {Herv{\'\i}as-Caimapo},
  {L{\'o}pez-Caniego}, {Martinez-Gonz{\'a}lez}, {Roman}, {Vielva}, {Wehus},
  {Achucarro}, {Ade}, {Allison}, {Ashdown}, {Ballardini}, {Banerji},
  {Bartlett}, {Bartolo}, {Baumann}, {Bersanelli}, {Bonato}, {Borrill},
  {Bouchet}, {Boulanger}, {Brinckmann}, {Bucher}, {Burigana}, {Buzzelli},
  {Cai}, {Calvo}, {Carvalho}, {Castellano}, {Challinor}, {Chluba}, {Clesse},
  {Colantoni}, {Coppolecchia}, {Crook}, {D'Alessandro}, {de Bernardis}, {de
  Gasperis}, {Diego}, {Di Valentino}, {Feeney}, {Ferraro}, {Finelli},
  {Forastieri}, {Galli}, {Genova-Santos}, {Gerbino}, {Gonz{\'a}lez-Nuevo},
  {Grandis}, {Greenslade}, {Hagstotz}, {Hanany}, {Handley},
  {Hernandez-Monteagudo}, {Hills}, {Hivon}, {Kiiveri}, {Kisner}, {Kitching},
  {Kunz}, {Kurki-Suonio}, {Lamagna}, {Lasenby}, {Lattanzi}, {Lesgourgues},
  {Lewis}, {Liguori}, {Lindholm}, {Luzzi}, {Maffei}, {Martins}, {Masi},
  {Matarrese}, {McCarthy}, {Melin}, {Melchiorri}, {Molinari}, {Monfardini},
  {Natoli}, {Negrello}, {Notari}, {Paiella}, {Paoletti}, {Patanchon}, {Piat},
  {Pisano}, {Polastri}, {Polenta}, {Pollo}, {Poulin}, {Quartin},
  {Rubino-Martin}, {Salvati}, {Tartari}, {Tomasi}, {Tramonte}, {Trappe},
  {Trombetti}, {Tucker}, {Valiviita}, {Van de Weijgaert}, {van Tent}, {Vennin},
  {Vittorio}, {Young}, \& {Zannoni}}]{2018JCAP...04..023R}
{Remazeilles}, M., {Banday}, A.~J., {Baccigalupi}, C., {et~al.}
  2018{\natexlab{a}}, \jcap, 2018, 023

\bibitem[{{Remazeilles} {et~al.}(2011){Remazeilles}, {Delabrouille}, \&
  {Cardoso}}]{2011MNRAS.410.2481R}
{Remazeilles}, M., {Delabrouille}, J., \& {Cardoso}, J.-F. 2011, \mnras, 410,
  2481

\bibitem[{{Remazeilles} {et~al.}(2018{\natexlab{b}}){Remazeilles}, {Dickinson},
  {Eriksen}, \& {Wehus}}]{2018MNRAS.474.3889R}
{Remazeilles}, M., {Dickinson}, C., {Eriksen}, H.~K., \& {Wehus}, I.~K.
  2018{\natexlab{b}}, \mnras, 474, 3889

\bibitem[{{Remazeilles} {et~al.}(2016){Remazeilles}, {Dickinson}, {Eriksen}, \&
  {Wehus}}]{2016MNRAS.458.2032R}
{Remazeilles}, M., {Dickinson}, C., {Eriksen}, H.~K.~K., \& {Wehus}, I.~K.
  2016, \mnras, 458, 2032

\bibitem[{{Remazeilles} {et~al.}(2021){Remazeilles}, {Rotti}, \&
  {Chluba}}]{2021MNRAS.503.2478R}
{Remazeilles}, M., {Rotti}, A., \& {Chluba}, J. 2021, \mnras, 503, 2478

\bibitem[{{Seljak}(1997)}]{1997ApJ...482....6S}
{Seljak}, U. 1997, \apj, 482, 6

\bibitem[{{Seljak} \& {Zaldarriaga}(1997)}]{1997PhRvL..78.2054S}
{Seljak}, U. \& {Zaldarriaga}, M. 1997, \prl, 78, 2054

\bibitem[{{Smith} \& {Zaldarriaga}(2007)}]{2007PhRvD..76d3001S}
{Smith}, K.~M. \& {Zaldarriaga}, M. 2007, \prd, 76, 043001

\bibitem[{{SO Collaboration}(2019)}]{2019JCAP...02..056A}
{SO Collaboration}. 2019, J. Cosmol. Astropart. Phys., 2019, 056

\bibitem[{{Starobinski{\v{i}}}(1979)}]{1979JETPL..30..682S}
{Starobinski{\v{i}}}, A.~A. 1979, Soviet Journal of Experimental and
  Theoretical Physics Letters, 30, 682

\bibitem[{{Stompor} {et~al.}(2016){Stompor}, {Errard}, \&
  {Poletti}}]{2016PhRvD..94h3526S}
{Stompor}, R., {Errard}, J., \& {Poletti}, D. 2016, \prd, 94, 083526

\bibitem[{{Stompor} {et~al.}(2009){Stompor}, {Leach}, {Stivoli}, \&
  {Baccigalupi}}]{Stompor2009}
{Stompor}, R., {Leach}, S., {Stivoli}, F., \& {Baccigalupi}, C. 2009, \mnras,
  392, 216

\bibitem[{{Thorne} {et~al.}(2019){Thorne}, {Dunkley}, {Alonso}, {Abitbol},
  {Errard}, {Hill}, {Keating}, {Teply}, \& {Wollack}}]{2019arXiv190508888T}
{Thorne}, B., {Dunkley}, J., {Alonso}, D., {et~al.} 2019, arXiv e-prints,
  arXiv:1905.08888

\bibitem[{{Thorne} {et~al.}(2017){Thorne}, {Dunkley}, {Alonso}, \&
  {N{\ae}ss}}]{2017MNRAS.469.2821T}
{Thorne}, B., {Dunkley}, J., {Alonso}, D., \& {N{\ae}ss}, S. 2017, \mnras, 469,
  2821

\bibitem[{{Vacher} {et~al.}(2022){Vacher}, {Aumont}, {Montier}, {Azzoni},
  {Boulanger}, \& {Remazeilles}}]{2022A&A...660A.111V}
{Vacher}, L., {Aumont}, J., {Montier}, L., {et~al.} 2022, \aap, 660, A111

\bibitem[{{Virtanen} {et~al.}(2020){Virtanen}, {Gommers}, {Oliphant},
  {Haberland}, {Reddy}, {Cournapeau}, {Burovski}, {Peterson}, {Weckesser},
  {Bright}, {van der Walt}, {Brett}, {Wilson}, {Millman}, {Mayorov}, {Nelson},
  {Jones}, {Kern}, {Larson}, {Carey}, {Polat}, {Feng}, {Moore}, {VanderPlas},
  {Laxalde}, {Perktold}, {Cimrman}, {Henriksen}, {Quintero}, {Harris},
  {Archibald}, {Ribeiro}, {Pedregosa}, {van Mulbregt}, \& {SciPy 1. 0
  Contributors}}]{2020NatMe..17..261V}
{Virtanen}, P., {Gommers}, R., {Oliphant}, T.~E., {et~al.} 2020, Nature
  Methods, 17, 261

\bibitem[{{Wagenmakers} \& {Farrell}(2004)}]{15117008}
{Wagenmakers}, E. \& {Farrell}, S. 2004, Psychon Bull Rev., 11, 192

\bibitem[{{Zaldarriaga} \& {Seljak}(1997)}]{1997PhRvD..55.1830Z}
{Zaldarriaga}, M. \& {Seljak}, U. 1997, \prd, 55, 1830

\bibitem[{{Zaldarriaga} \& {Seljak}(1998)}]{1998PhRvD..58b3003Z}
{Zaldarriaga}, M. \& {Seljak}, U. 1998, \prd, 58, 023003

\bibitem[{Zonca {et~al.}(2019)Zonca, Singer, Lenz, Reinecke, Rosset, Hivon, \&
  Gorski}]{Zonca2019}
Zonca, A., Singer, L., Lenz, D., {et~al.} 2019, Journal of Open Source
  Software, 4, 1298

\end{thebibliography}

\appendix

\section{Validation of power spectra and posteriors}\label{app:validation}

\begin{figure}
    \centering
     \includegraphics[width=\columnwidth]{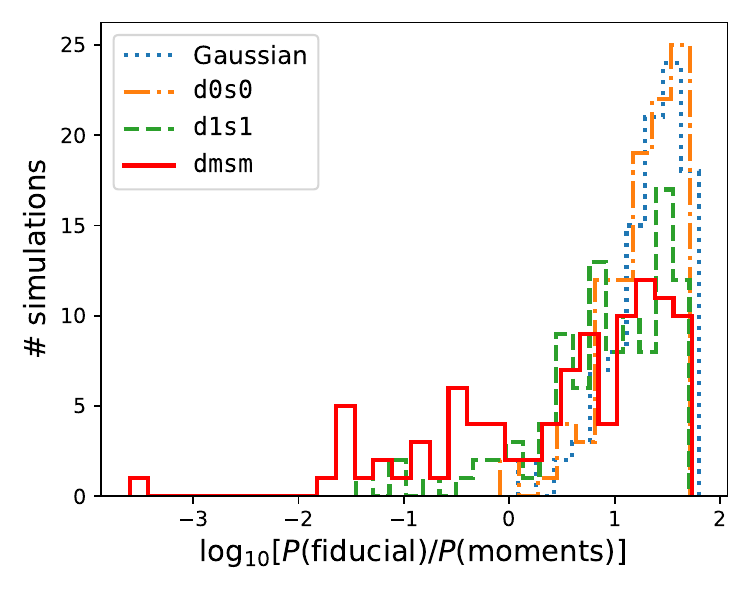}
    \caption{Empirical distribution of the AIC-based relative model odds between the $C_\ell$-fiducial and the $C_\ell$-moments model from 100 simulations. Same as Fig.~\ref{fig:model_odds_d10s5} but only considering the four less complex Galactic foreground templates. Note the different $x$ scale. We do not see strong preference for $C_\ell$-moments (indicated by large negative values) in any of the cases considered.} \label{fig:model_odds}
\end{figure}

The goodness of fit of the power spectrum likelihood can be assessed by the well-known minimum-$\chi^2$ statistic $\hat{\chi}^2_{\rm min} = (\mathbf{d}-\mathbf{\hat{t}})^T {\sf C}^{-1}(\mathbf{d}-\mathbf{\hat{t}})$, with $\mathbf{d}$, $\mathbf{\hat{t}}$, and ${\sf C}$ denoting data, best-fit theory model, and covariance, respectively. We perform this validation for pipeline A only, since it includes the foreground model at the likelihood level. Under the hypothesis of Gaussian data, $\hat{\chi}^2_{\rm min}$ is expected to follow a $\chi^2$ distribution with $N-P=558$ degrees of freedom \citep{pearson1900,Maltoni:2003cu}. When using Gaussian foregrounds, we know that our data are sufficiently Gaussian and the covariance is exact within the MCMC noise level. We compute the $\hat{\chi}^2_{\rm min}$ for 100 simulated data realizations containing standard cosmology ($r=0$, $A_{\rm lens}=1$), inhomogeneous noise in the goal-optimistic scenario and in four foreground cases (Gaussian, \dzsz{}, \doso{}, \dmsm{}), considering both the $C_\ell$-fiducial and the $C_\ell$-moments model. The empirical distributions match the theoretical expectation in all cases, showing that the covariance matrix is appropriate, even for non-Gaussian foreground input templates.

We also assess, for each simulation seed, which foreground model is preferred by the simulated data, using the \textit{Akaike Information Criterion} \citep[AIC, ][]{1100705}. We compute the difference $\Delta{\rm AIC}=2\Delta k + \hat{\chi}^2_{\rm min}({\rm moments})-\hat{\chi}^2_{\rm min} ({\rm fiducial})$, where $\Delta k=4$ is the number of excess parameters of the $C_\ell$-moments over $C_\ell$-fiducial. Following \cite{15117008}, the number $\exp(\Delta{\rm AIC}/2)$ can be interpreted as the relative model odds $P(C_\ell$-fiducial$)/P(C_\ell$-moments$)$.

The results are shown in Fig.~\ref{fig:model_odds}. The AIC test detects no clear preference for $C_\ell$-moments over $C_\ell$-fiducial. Among 100 simulations containing \dmsm{} foregrounds, 30 have model odds below 1, and only a single simulation below $10^{-2}$. In case of the \dmsm{} simulation set, this might come as a surprise, considering that the $C_\ell$-moments model allows to mitigate a 1-2$\sigma$ bias. We conclude that one must be careful when interpreting $<2\sigma$ detections in marginal posterior distributions, as they may be difficult to distinguish from subdominant residual foreground contamination using standard model-comparison techniques. As shown in Sect.~\ref{sec:d10s5}, this situation changes in the presence of input foregrounds that induce a large bias.

Lastly, we tested the robustness of the statistical results on $r$ quoted in Sect.~\ref{sec:results}. Using pipelines A, A+moments, and B, we calculated the mean and maximum a-posteriori (MAP) value of the marginal $r$ posterior for each of 500 simulations containing CMB realized with the fiducial cosmology ($r=0$, $A_{\rm lens}=1$), \dmsm{} foregrounds, and inhomogeneous noise in the baseline-optimistic scenario. We computed the average and standard deviation over the 500 simulations, and repeated the procedure for the lensing amplitude $A_{\rm lens}$. Table~\ref{tab:posterior_validataion} shows the results. We find that for both parameters and all three pipelines, the sample average of the mean and the MAP agree at the $0.1\sigma$ level. We also find consistency between the standard deviation of the marginal posteriors averaged over 500 simulations, the sample scatter of the marginal posterior mean values, and the sample scatter of the MAP values computed from 500 simulations. Table~\ref{tab:posterior_validataion} also lists the Gaussian error on the average mean and MAP values, corresponding to the sample scatter divided by the square root of the number of simulations, $\sqrt{500}$. While this test would result in a bias on $r$ for all component separation pipelines (yet below $2\sigma$ in the case of the A + moments pipeline), passing it is far beyond the requirements set by the statistical sensitivity of SO.

\begin{table*}
\caption{Posterior statistics on $r$ and $A_{\rm lens}-1$ from 500 simulations. We assume \dmsm{} foregrounds and baseline-optimistic noise, as inferred by the three nominal component separation pipelines. No delensing or primordial gravitational waves are assumed.} 
\label{tab:posterior_validataion}
    \centering
    \begin{tabular}{ccccccccc}
    \hline
    \hline
    \noalign{\smallskip}
    & & \multicolumn{3}{c}{posterior mean} & \multicolumn{3}{c}{posterior maximum} & posterior \\
    Pipeline & parameter & mean & $\sigma$ & $\Delta$ & mean & $\sigma$ & $\Delta$ & standard deviation \\
    \noalign{\smallskip}
    \hline
    \noalign{\smallskip}
    A & \begin{tabular}{@{}c@{}}$r\times10^3$ \\ $(A_{\rm lens}-1)\times10^2$ \end{tabular} 
    & \begin{tabular}{@{}c@{}}$3.9\pm 0.1$ \\ $-0.6\pm 0.1$ \end{tabular}
    & \begin{tabular}{@{}c@{}}$2.6$ \\ $3.3$ \end{tabular} 
    & \begin{tabular}{@{}c@{}}$+1.5$ \\ $-0.2$ \end{tabular} 
    & \begin{tabular}{@{}c@{}}$4.0\pm0.1$ \\ $-0.7\pm0.1$ \end{tabular} 
    & \begin{tabular}{@{}c@{}}$2.6$ \\ $3.3$ \end{tabular} 
    & \begin{tabular}{@{}c@{}}$+1.6$ \\ $-0.2$ \end{tabular} 
    & \begin{tabular}{@{}c@{}}$2.6$ \\ $3.5$ \end{tabular} \\
    \noalign{\smallskip}
    \hline
    \noalign{\smallskip}
    A+moments & \begin{tabular}{@{}c@{}}$r\times10^3$ \\ $(A_{\rm lens}-1)\times10^2$ \end{tabular}
    & \begin{tabular}{@{}c@{}}$0.3\pm 0.1$ \\ $0.0\pm 0.1$ \end{tabular}
    & \begin{tabular}{@{}c@{}}$4.0$ \\ $3.4$ \end{tabular} 
    & \begin{tabular}{@{}c@{}}$+0.1$ \\ $0.0$ \end{tabular} 
    & \begin{tabular}{@{}c@{}}$-0.1\pm0.2$ \\ $0.0\pm0.1$ \end{tabular} 
    & \begin{tabular}{@{}c@{}}$4.0$ \\ $3.4$ \end{tabular} 
    & \begin{tabular}{@{}c@{}}$0.0$ \\ $0.0$ \end{tabular} 
    & \begin{tabular}{@{}c@{}}$3.5$ \\ $3.6$ \end{tabular} \\
    \noalign{\smallskip}
    \hline
    \noalign{\smallskip}
    B & \begin{tabular}{@{}c@{}}$r\times10^3$ \\ $(A_{\rm lens}-1)\times10^2$ \end{tabular} 
    & \begin{tabular}{@{}c@{}}$2.5\pm 0.1$ \\ $6.2\pm 0.1$ \end{tabular} 
    & \begin{tabular}{@{}c@{}}$2.7$ \\ $3.1$ \end{tabular} 
    & \begin{tabular}{@{}c@{}}$+0.9$ \\ $+2.0$ \end{tabular} 
    & \begin{tabular}{@{}c@{}}$2.5\pm0.1$ \\ $6.2\pm0.1$ \end{tabular} 
    & \begin{tabular}{@{}c@{}}$2.7$ \\ $3.1$ \end{tabular} 
    & \begin{tabular}{@{}c@{}}$+0.9$ \\ $+2.0$ \end{tabular} 
    & \begin{tabular}{@{}c@{}}$2.7$ \\ $3.2$ \end{tabular} \\
    \noalign{\smallskip}
    \hline
    \noalign{\smallskip}
    \hline
    \end{tabular}
\end{table*}

\normalsize

\section{Bias on $r$ for Gaussian foregrounds and pipeline B} \label{sec:nilc_gaussian_bias}

\begin{figure*}
    \centering
    \includegraphics[width=0.49\textwidth]{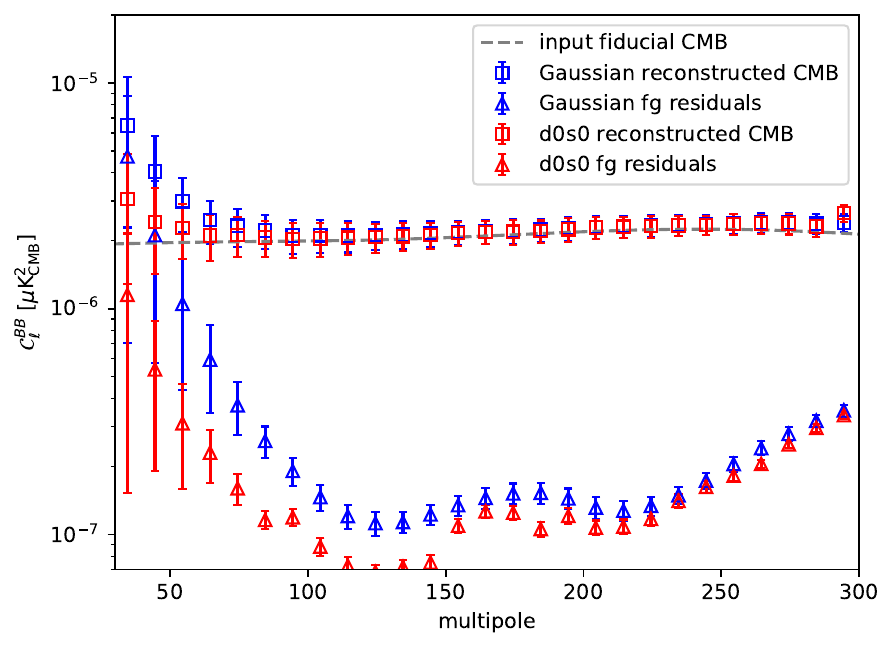}
    \includegraphics[width=0.49\textwidth]{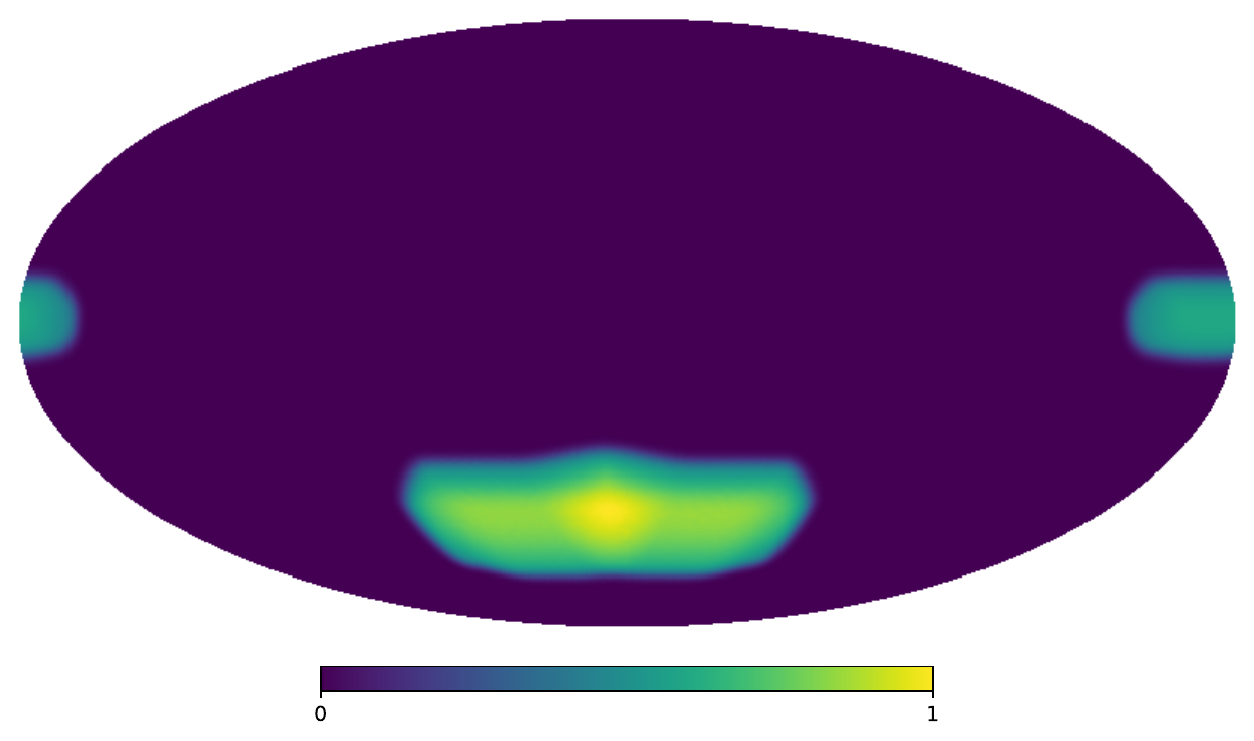}
    \caption{Foreground bias and custom analysis mask related to needlet-based component separation. \textit{Left panel:} Foreground residuals in the reconstructed CMB power spectra in the inhomogeneous goal-pessimistic noise scenario, with a fiducial input cosmology of $r=0$, $A_{\rm lens}=1$, and Gaussian and \dzsz{} foregrounds. We show the reconstructed CMB and foreground power spectrum residuals $C_{\ell}^{BB}$. The residual is calculated by mixing the pure foreground maps with the NILC weights. \textit{Right panel:} constrained analysis mask used by pipeline B in the case of Gaussian foregrounds, built from the fiducial mask shown in Fig.~\ref{fig:mask}. \label{fig:pipelineB_foreground_bias}}
\end{figure*}

In the results of pipeline B, we note a consistent bias on $r$ for simulations that include Gaussian foregrounds in combination with inhomogeneous noise. This bias seems unreasonable, considering  the results that use more complex foregrounds, such as \dzsz{}, but the same noise and CMB simulations, lead to considerably less bias. The cosmological bias is directly caused by a bias at the large angular scales of the $BB$ power spectrum, which is easily visible when plotting the spectra. We confirm that this bias corresponds to foreground bias. In Fig.~\ref{fig:pipelineB_foreground_bias} (left panel), we show the reconstructed CMB $C_{\ell}^{BB}$ spectrum for the standard cosmology ($r=0$, $A_{\rm lens}=1$) with inhomogeneous goal-pessimistic noise, for both the Gaussian (blue squares) and \dzsz{} (red squares) foregrounds. The marker and error bar shows the mean and 1-$\sigma$ standard deviation across the 500 simulations. The excess bias at large scales is evident as the reconstructed CMB clearly surpasses the fiducial $BB$ spectrum. We can calculate the exact foreground bias present on each reconstructed CMB, by mixing the pure foregrounds maps with the same NILC weights (Eq.~\ref{eq:nilc_weights}), and taking the power spectra of that foreground bias reconstruction. This is shown as the blue triangles for Gaussian and red triangles for \dzsz{} foregrounds. In the Gaussian case, we see that the large-scale bias is almost entirely caused by foregrounds.

We know this bias is directly related to the NILC weights that mix the frequency maps (Eq.~\ref{eq:nilc_weights}). The weights are calculated directly from the frequency-frequency covariance matrix, which in turn is calculated from the input frequency maps themselves. If, for instance, we wish to test if the weights calculated with the simulations including Gaussian foregrounds are incorrect, we can do the cross check by using other weights to mix those same frequency maps. We take the NILC weights calculated for the simulations with \dzsz{} foregrounds (shown in Fig.~\ref{fig:pipelineB_foreground_bias}) and use them to mix the frequency maps that include the Gaussian foregrounds. In this case, the mean best fit is $r=-0.0001 \pm 0.0026$, which is comparable to $r = -0.0005 \pm 0.0028$ for the simulations including \dzsz{} foregrounds (shown in Table~\ref{tab:r_fiducial_results}). 

\begin{figure*}
    \centering
    \includegraphics[width=1.0\textwidth]{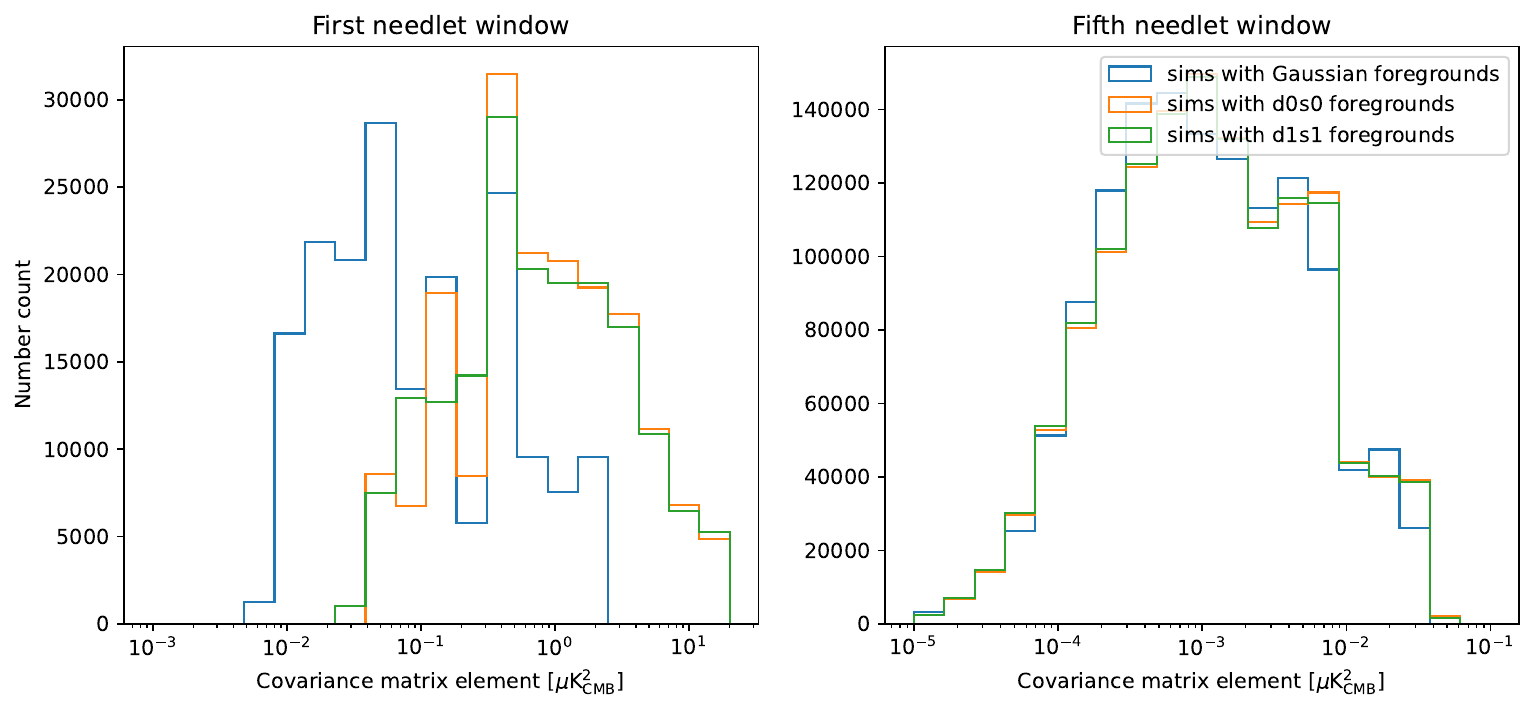}
    \caption{Distribution of covariance matrix elements used to build the ILC weights (see Eq.~\ref{eq:nilc_weights}) for pipeline B. We show values for a single simulation seed used for B-mode reconstruction, and include all cross-frequency combinations. We assume inhomogeneous goal-pessimistic noise, with a fiducial ($r=0$, $A_{\rm lens}=1$) CMB, with Gaussian, \dzsz{}, and \doso{} foregrounds. \textit{Left panel:} first needlet window (largest scales), \textit{right panel}: fifth needlet window (smallest scales). \label{fig:covariance}}
\end{figure*}

Any incorrect weights must originate from the NILC covariance matrix. The combination of directly calculating the covariance matrix over maps that have inhomogeneous noise and Gaussian foregrounds creates the observed bias. We can alleviate this problem by spatially constraining the hits mask (Fig.~\ref{fig:mask}) to a more homogeneous area, where the noise map realization will also be more homogeneous. The mask we use to do this is shown in Fig.~\ref{fig:pipelineB_foreground_bias} (right panel). This custom mask has a smaller sky fraction than the fiducial analysis mask, which increases the statistical uncertainty on the inferred value of $r$. The measurements using this more constrained and more homogeneous mask are marked with a $^{\dagger}$ in Table~\ref{tab:r_fiducial_results}.

\end{document}